\documentclass[useAMS,usenatbib]{mn2e}
\usepackage{graphicx}
\newcommand{\swift}{SWIFT J1753.5$-$0127}
\newcommand{\xte}{XTE J1118+480}
\newcommand{\gx}{GX 339$-$4}

\title[\swift: multiwavelength observations and
  reactivation]{Multiwavelength spectral and high time resolution 
  observations of \swift: new activity?}
\author[Durant et al.]{M. Durant$^{1}$\thanks{E-mail:
durant@iac.es}, P. Gandhi$^2$, T. Shahbaz$^1$, H. H. Peralta$^{1}$ and
  V. S. Dhillon$^3$\\
$^1$ Instituto de Astrof\'isica de Canarias, La Laguna, E38205 Tenerife,
  Spain\\
$^2$ RIKEN Institute of Physical and Chemical Research, 2-1 Hirosawa,
  Wakoshi, Saitama, Japan\\
$^3$ Department of Physics and Astronomy, University of Sheffield,
  Sheffield S3 7RH, UK
}
\begin{document}

\date{~}

\pagerange{\pageref{firstpage}--\pageref{lastpage}} \pubyear{2008}

\maketitle

\label{firstpage}

\begin{abstract}
We have conducted an extensive observational campaign of
\swift\ during June 2007 after its bright outburst episode in
2005. We have performed multi-band optical photometry, optical
spectroscopy, X-ray spectroscopy and timing and ULTRACAM optical
photometry simultaneously in three bands. Both the optical spectrum
and the X-ray spectrum, along with enhanced brightness in broad-band
photometry point to recent increased activity.  We analyze the
different spectral regions, finding a { smooth optical continuum} with a
remarkable lack of lines and a very blue component modulated with a period of
3.2\,hr and a hard power-law X-ray spectrum.  Both the X-ray and
optical power spectra are flat at low frequencies up to the 0.1\,Hz
(10\,s) range, then decreasing roughly as a power law consistent with
flickering. Furthermore, the optical data show quasi-periodic
oscillations (QPOs) near 0.08\,Hz (13\,s). Together with a dynamical
and auto-correlation analysis of the light curves we attempt to
construct a complete physical picture of this intriguing system.
\end{abstract}


\section{Introduction}
Soft X-ray Transients (SXRTs) are a subset of the  low mass X-ray binaries
(LMXBs) which are 
thought to contain a black hole primary, and undergo periodic brightening
episodes lasting a few months, with recurrence times of a few years (Tanaka \&
Shibazaki, 1996). The outburst
episodes typically coincide with spectral and timing changes, for
example the X-ray {\em high/soft} state (thermal emission when accretion is
high) and the X-ray {\em low/hard} state (coronal/jet or optically
thin disc emission when accretion is low; review in Remillard \&
 McClintock, 2006). Such changes are
attributed to varying contributions from the various emission
mechanisms e.g., dense disc, jet, low density zone (e.g., van der
Klis, 2006). It turns out,
however, that such a classification may be 
simplistic, as some systems have been seen to go into outbursts
without changing their X-ray state. It is a challenge to understand, in
these cases, what indeed is changing (e.g., Meyer-Hofmeister, 2004).

Two main models exist as possible explanations for the low/hard state
in accreting binaries. The first and oldest is the Advection-Dominated
Accretion Flow (ADAF, e.g., Czerny et al. 2000) model, in which the
accretion disc is truncated at a 
large inner radius, and the relatively inefficient ADAF channels
matter to the compact object and to any jets; this region is
low-density and responsible for the high-energy emission. In a typical
outburst cycle, the accretion disc gains enough mass to overwhelm the
ADAF, and fast accretion progresses until the disc is depleted,
whereupon the ADAF reforms from small radii to large. The contender
model, the Accretion Disc Corona (ADC; e.g., Malzac, 2007) is not
dissimilar in its 
effects, but the mechanism is subtly different. A hot corona of
energetic particles may form above and below an accretion disc,
and most of the accretion power is carried by strong magnetic
fields. When the fields in 
the corona are strong, the emission from the corona is strong, and a
higher proportion of in-falling material is channelled into the
jets. When the mass-flow rate increases beyond a critical level,
however, enough soft X-ray photons are emittied by the disc to
efficiently cool the corona, causing it to condense onto the
disc, halting the chanelling of energy and particles away from the
core region, which then dominates the emitted spectrum with thermal
radiation from small radii. Note that this model was developed
initially fpr sources where the central accretion disc is obscured by the
outer torus of material, making the ADC relatively more important.

In general, it may be
assumed that both processes, with their different physical mechanisms,
occur in systems either simultaneously or episodically. The most
obvious difference between the two is that for the ADC, the accretion
disc extends down to very small radii, to the last stable orbit in the
case of a black-hole system. The
transitions between accretion modes and feedback in the systems is
very poorly understood, however. It seems not to be purely a function
of accretion rate, and there certainly appears to be hysteresis in all
the transient systems, as typically depicted by their path through the
X-ray hardness-luminosity diagram (Homan \& Belloni, 2005).

One object which has received a lot of attention in recent times,
being an out-lier in the SXRT population, is \xte\ (Chaty et
al. 2003). This system was 
thoroughly observed through outburst to quiescence in X-rays, UV,
optical and radio (Hynes at al. 2003), and showed interesting breaks
and possible QPOs in its periodograms (Shahbaz et al. 2005). Being
high above the Galactic plane, it has been 
suggested that \xte\ belongs to a new population of black hole
binaries, with the question of how they can appear in the halo still
open (McClintock et al. 2001). \xte\ is one of the few
systems which does not 
follow the typical path for an SXRT through X-ray hardness/luminosity
space: it stayed in the low/hard state throughout. It is also the only
system for which a high signal-to.noise optical/X-ray
cross-correlation has been calculated, which showed an unexpected
``precursor'' anti-correlation with optical leading X-rays (and a
stronger positive resonse; Kanbach et al. 2001).

The SXRT \swift\ is an X-ray transient system which has been of
great interest recently following its outburst episode and
detailed observations with the {\em SWIFT} satellite. First discovered
by the {\em SWIFT}/BAT (Burst Alert Telescope; Palmer et al., 2005)
in 2005, pointed $\gamma$-ray, X-ray, UV, 
optical and radio observations all detected a new bright source at
this location (Morris et al., 2005; Still et al., 2005; Halpern et
al., 2005; Fender et al., 2005). {\em Swift}/XRT (X-ray Telescope;
Burrows et al. 2005), and {\em RXTE} (Rossi X-ray Timing Explorer;
Jahoda et al. 1996)
observations detected the existence of 
a strong 0.6\,Hz quasi-periodic oscillation (QPO; Morgan et al. 2005;
Ramadevi \& Seetha, 2005), persistent for some time
after the outburst episode. This QPO has so far only been seen in
X-ray observations. Zhang et al. (2007) presented the evolution of the
QPO frequency with time after the outburst, and its relationship to
X-ray hardness. With a high Galactic latitude, comparisons with
\xte\ promise to be very interesting.

Summarizing the above works very briefly, after a fast initial rise in
X-ray flux, the source returned to its previous level, almost
undetectable by the RXTE All-sky Monitor, in a few months (rise time
of 4.5 days, decay of an exponential time-scale initially $\sim$32\,d,
later $\sim$100\,d, peak flux
$\sim$0.2\,crab in the RXTE/PCA or SWIFT/XRT band). Throughout,
it remained hard in X- and $\gamma$-rays (a rising spectrum in $\nu
f_\nu$ out to $\sim$200\,keV), suggesting a hot Comptonizing
corona (ADC). The optical counterpart also faded, but the fade slowed, and
it remains relatively bright and blue. H$_\alpha$ and He lines were
originally observed, double-peaked, but disappeared after the initial
outburst (Torres, 2005). The accretion disc inclination is thus
non-negligible, but neither is it extreme ($\geq80$\degr), since eclipses
are never observed. The H$_\alpha$ line was not particularly strong
(equivalent width $\sim3$\,\AA) but very broad
(FWHM$\sim$2000\,km\,s$^{-1}$). From the Na{\tt I} doublet and UV
observations, the 
extinction equivalent to $N_H \sim 2\times 10^{21}$\,cm$^{-2}$
and a distance of $\sim 6$\,kpc have been estimated. The radio spectrum
was variable and consistent with a flat power law ($f_\nu \sim \nu^0$)
 which lies underneath the optical points, if extrapolated (Fender et
al. 2005). This suggests that, initially at least, the optical
emission did not appear like a typical synchrotron spectrum.

Cadolle Bel et al. (2007; CB07 from hereon) performed follow-up simultaneous
multi-wavelength observations. They identify \swift\ as a likely
Black-Hole Candidate, which stayed in a low/hard state throughout its
outburst and gradual fade. They also detected a QPO in their X-ray
data, but weaker, and at the somewhat lower frequency of 0.24\,Hz. Such a
reduction in the frequency is consistent with what would be expected
for the inner radius of an accretion disc, which expands during the
decaying lifetime of the outburst in the ADAF model. The INTEGRAL
count-rate during their 
observation was constant at 43\,cts\,s$^{-1}$ ($\sim205$\,mCrab
between 20--320\,keV), with a hardness ratio
$HR=f_{20-40\,\textrm{keV}}/f_{40-80\,\textrm{keV}}\sim0.75$. They
also found a flat radio spectrum with fluxes of the order
0.65\,mJy.

Miller et al. (2006a) observed \swift\ some months after the outburst
episode above with XMM-Newton and RXTE, when the source was assumed to have
reached quiescence. Their spectral modelling indicated
 a prominent accretion disc was required by
the X-ray spectrum with high significance. This disc is cool ($kT =
0.2$\,keV), and extends to very small radii near the last inner stable
orbit. This model would pose problems for models of low-level
accretion involving an ADAF region near the central source, and
complicate jet-formation scenarios. An ADAF model is commonly invoked
to explain the existence of a jet in the low/hard state and its
absence in the high/soft state, in systems where both states have been
observed (e.g., Fender, 2006).

We have conducted a comprehensive observational campaign of \swift,
including optical photomety in different filters on five successive
nights, with $\sim1$\,min time resolution, contemporary X-ray
observations, optical spectroscopy and high time-resolution optical
observations with ULTRACAM.  Here we present the results of this
campaign, particularly timing analyses of the data-sets and
comparisons between them. In the companion paper, Durant et al.
(2008) analyze the cross-correlation of the X-ray and optical
light curves from the simultaneous high-speed observations, finding
suprisingly that the optical precedes the X-rays with a broad
anticorrelation peak, followed by a weak positive response for the
optical lagging the X-rays.  The only contemporaneous radio
observation of \swift\ we are aware of were taken by Soleri et
al. (2008), who did not detect the source with the Westerbrook
Synthesis Radio Telescope in July 2007, establishing a 3$\sigma$
limiting flux of 1.1\,mJy at 5\,GHz and 8\,Ghz. Zurita et al. (2008)
also observed this system in the optical band, but over a much longer
time base-line (several months) and they discuss the long-term optical
brightness trend and newly discovered $\sim 3.2$\,hr superhump/orbital
period.

In the following section we describe the various observations and
processing. In Section 3, we give the results of this, and further
analysis comparing the different data, including timing and spectral
analyses. In Section 4, we discuss the results and attempt to draw
physical conclusions from them.

\section[]{Observations}
Table \ref{obslog} lists all the observations of \swift\ that are
analyzed here. Of the observations listed there, only the WHT/ISIS
spectrum was {\em not} obtained during the same week of 2007 (it is
from one year earlier), whereas the radio limits of Soleri et
al. (2008) were approximately contemporary with our observations.

Further to our own observations below, we retrieved the Rossi All Sky
Monitor (AMS; Levine et al., 1996) daily average count rates for
\swift, as well as regular monitoring observations by the gamma-ray
satellite INTEGRAL's 
wide-field lower energy-band imaging instrument ISGRI (Integral Soft Gamma-ray
Instrument; Lebrun et al. 2003). These are shown in Figure
\ref{ASM}. The ASM observes the 
whole sky, and INTEGRAL surveys the Galactic Centre regularly, and all
interesting bright objects are automatically measured. These data are
publicly available (the reduction is not described here, see the
references above).

\subsection{X-ray}

\swift\ was observed for 53.6\,min each on 11 Jun 2007 and 13 Jun
2007 with the {\em Rossi X-ray Timing Explorer} (RXTE). 
 RXTE comprises three
instruments: the Proportional Counting Array (PCA; Jahoda et al. 1996)
for soft-band pointed observations with large effective area, the
High-Energy X-ray Timing Experiment (HEXTE; Rothschild et al. 1998)
for energies up to 200\,keV and the All-Sky 
Monitor (ASM, see above), with a very large effective field of view. None are
imaging instruments, but provide high temporal resolution. For the
first observation, only two of the PCA units were functional, and
in the second observation, three. We use only the PCA light curves in
our timing analysis, since the count rates are so much higher than the
high-energy HEXTE count rates. The latter we only use in the spectral
fitting. 

Events were processed through the standard pipeline for both the PCA
and HEXTE data-sets. Standard Good Time Intervals were applied and we
selected  all events flagged as good for further
analysis. We generated light curves by binning the events on a regular
grid; uncertainties on each sample are dominated by photon counting
noise. 

When a fluxed spectrum is generated from the data, using the
instrument response functions, the HEXTE and PCA parts of the spectrum
do not appear to meet: there seems to be a 15\% discrepancy, 
which corresponds to the effect of dead time. We corrected for this
and performed our spectral analysis on these data.

\subsection{Optical Photometry}
We observed \swift\ with the 2.2\,m Nordic Optical Telescope (NOT; 2007
Jun 16, 17) and the 1.5\,m Mercator (2007 Jun 18, 19, 20) telescopes
at Roque de los Muchachos Observatory, 
La Palma, Spain. In each case, we performed CCD-based imaging with
exposure times of 60\,s. With the NOT we acquired images in the $U$ and $B$
filters, and with the Mercator in the $BVR$ filters. Standards were
imaged with the NOT in the $U$ and $B$ filters, and the Mercator
photometry was calibrated via images of both the science field and
standards fields with the IAC80 telescope at the Teide Observatory,
Tenerife. The weather was generally good, especially for the NOT run,
with some variability and thin cloud during the Mercator run.

Images were in every case first bias-subtracted, then flat-fielded
using drift images of the twilight sky. Note that the NOT/Alfosc
images were affected by unrepeatable electronic pick-up, and that the
photometry was therefore somewhat degraded. Photometry was measured in small
apertures relative to a bright local reference in the field. For this,
we used the ULTRACAM pipeline, as below with the ULTRACAM data. The
local reference was in each case calibrated relative to the Landolt
standard list (Landolt, 1992), correcting for a colour term for each filter
set. Having other stars in the field of similar brightness 
 to \swift\ enabled us to check that the analysis process was not
introducing any systematic signal into the light curves.
The uncertainty on each measurement is estimated from the photon
statistics of the counts within an aperture and the sky
background. Our measurements are source photon noise dominated.

\subsection{Optical Spectroscopy}
We obtained spectra across the optical range with the
Intermediate dispersion Spectrograph and Imaging System
(ISIS\footnote{\tt http://www.ing.iac.es/Astronomy/observing/manuals/
  html\_manuals/wht\_instr/isis\_hyper/isis\_hyper.html}) on the
4.2\,m William Herschel Telescope (WHT) at the Observatorio Roque 
de los Muchachos, La Palma, Spain. ISIS features twin spectrographs
optimized for the red and blue end of the optical, which can be used
simultaneously with the use of a dichroic in the main beam. We obtained
1.72\,\AA/pixel mean dispersion in the blue arm, and 1.65\,\AA/pixel
in the red, with a mean resolution $\sim3$\,\AA\ throughout, under good
conditions. The two spectra were wavelength calibrated and extracted
using standard {\tt iraf} tools, and co-added into master spectra. In
order not to lose any resolution when co-adding, each spectrum was
first re-sampled to 0.5\,\AA.

To flux and produce the final spectrum, we observed a spectral
standard at a similar airmass, and for both the object and the
standard spectrum, summed together the red and blue parts into a
single master spectrum (sampled at 0.5\,\AA). Since the dicroic
provides a sharp but finite cut-off, calculating the response function
for any one arm in the cut-off region is hard, but yet there are
enough photons in total at any given wavelength to find the overall
sensitivity. The final response curve we used to flux the data was a
heavily smoothed version of the ratio of the tabulated flux to the
measured counts for the standard.
These observations were done one year after the outburst of \swift, during
the faintest part of the ASM light curve since detection (Figure
\ref{ASM}). 

We also obtained a spectrum of \swift\ during our June observing
campaign with the Focal Reducer and low dispersion Spectrograph
(FORS2, Appenzeller et al. 1998) on the 8.2\,m Unit Telescope 1 (Antu)
of the VLT, Cerro Paranal, Chile. This observation was
simultaneous with one RXTE and one ULTRACAM 
observation. The spectral scale was 3.17\,\AA/pixel, and resolution
$\sim7$\,\AA. Conditions were 
mediocre, with variable cloud cover, transparency and seeing. Since
the whole spectral range is obtained in a single go, the problems
associated with matching two spectra (above) were not encountered
here, and the extraction was straight-forward.

\subsection{ULTRACAM}
In addition, \swift\ was
also observed with ULTRACAM, mounted on the VLT/3 (Melipal) telescope on the
nights of 2007 Jun 12 and 17, for 1.3\,hr and 0.5\,hr
respectively. ULTRACAM is an instrument employing dichroic beam 
splitters, frame-transfer CCDs and a GPS-based timing system in order
to be able to make simultaneous multi-wavelength optical light curves at very
high time resolution, up to 500\,Hz (Dhillon et al. 2007). We used
two small windows on each CCD (one for the 
source of interest, one for a local standard), with exposure times of
140\,ms for the 12th and 39\,ms for the 17th (and duty cycles of
142\,ms and  41\,ms respectively). The reason for the
difference in exposure times was thin cloud on the first of the two
nights, giving similar signal-to-noise per image.

The object was visible in every frame in the $r'$ and $g'$ 
bands, but not in the $u'$ band. For the latter, therefore, the source
could only be 
detected by co-adding many images, resulting in  reduced temporal
resolution but better signal-to-noise. On the first night, with
the poor conditions, the effect 
was particularly strong, and we do not attempt to analyze these
data. On the 17th, however, it was possible to get reasonable
measurements from averages of every 50 images. Thus it was not possible
to search for high-frequency variability in these data, but timescales
$\geq$4\,s were accessible.

Fluxes were extracted by aperture photometry with a variable aperture
size scaled to the FWHM of the reference star on each image. This
enables some optimization for signal-to-noise under variable
conditions. The optimal extraction method (Naylor, 1992) did not yield
appreciably different results, since that method is more applicable to the
faint, background-dominated regime.

The night of the 2007 June 12 was badly affected by 
transparency variations. In addition, the comparison star chosen was
of similar brightness to our source in $r'$ and significantly redder,
introducing additional uncertainty in $r'$ and even more in $g'$. This is
evidenced by the difference in the 
$r'$-band and $g'$-band light curves. Only for the 17th can we give
reliable average magnitudes, based on the calibrated zero-points of
the instrument. These are on the SDSS photometric system 
(systematic uncertainties here are $\pm$0.03--0.05\,mag, increasing to the
blue). By the transformations given in Jester et al.
(2005), the SDSS average magnitudes correspond to V$=$16.62(5),
B$=$16.92(7), consistent with the slow photometry above. The slow
photometry has a better absolute calibration.

\begin{table*}[h!]
\begin{center}
 \caption{Observation log of \swift}
 \label{obslog}
 \begin{tabular}{lccccc}
  \hline
Date (UT) & Type & Instrument & Filter(s) & Duration (min) \\
  \hline
2006-06-17 & Spectrum & WHT/ISIS & red, blue &160\\
2007-06-11 & X-ray & RXTE & & 54\\
2007-06-13 & Photometry & ULTRACAM & $u'g'r'$ & 78\\
2007-06-13 & X-ray & RXTE & & 54\\
2007-06-13 & Spectrum & VLT1/FORS2 & & 3\\
2007-06-16 & Photometry & NOT/Alfosc & $UB$ & 434\\
2007-06-17 & Photometry & NOT/Alfosc & $U$ & 420\\
2007-06-17 & Photometry & ULTRACAM & $u'g'r'$ & 30 \\
2007-06-18 & Photometry & Mercator/Merope & $BV$ & 411\\
2007-06-19 & Photometry & Mercator/Merope & $BVR$ & 290\\
2007-06-20 & Photometry & Mercator/Merope & $BVR$ & 432\\
  \hline 
 \end{tabular}
\end{center}
\end{table*}

\section{Analysis and Results}

\subsection{Long-term trend}
\begin{figure*}\includegraphics[width=0.8\hsize]{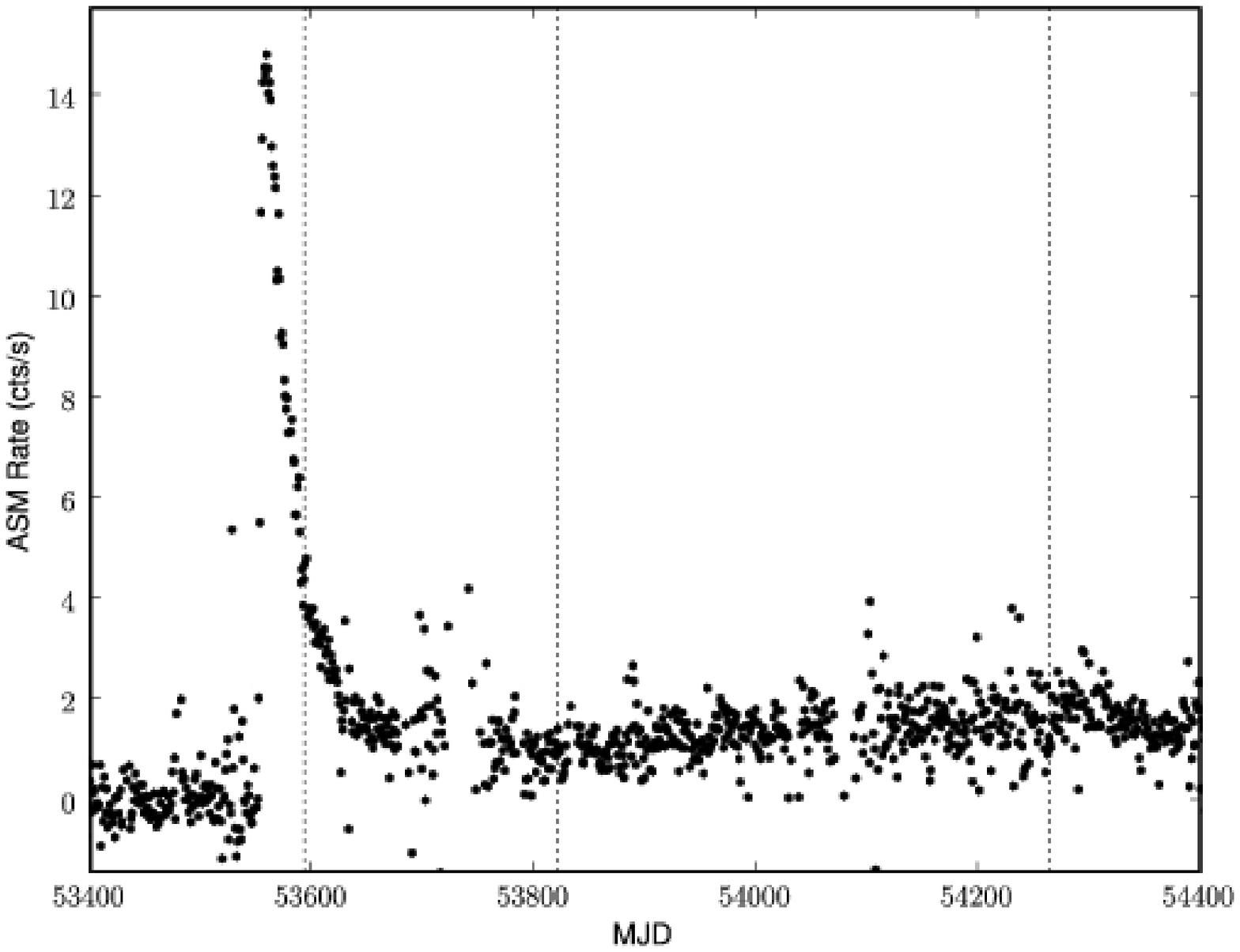}
\includegraphics[width=0.8\hsize]{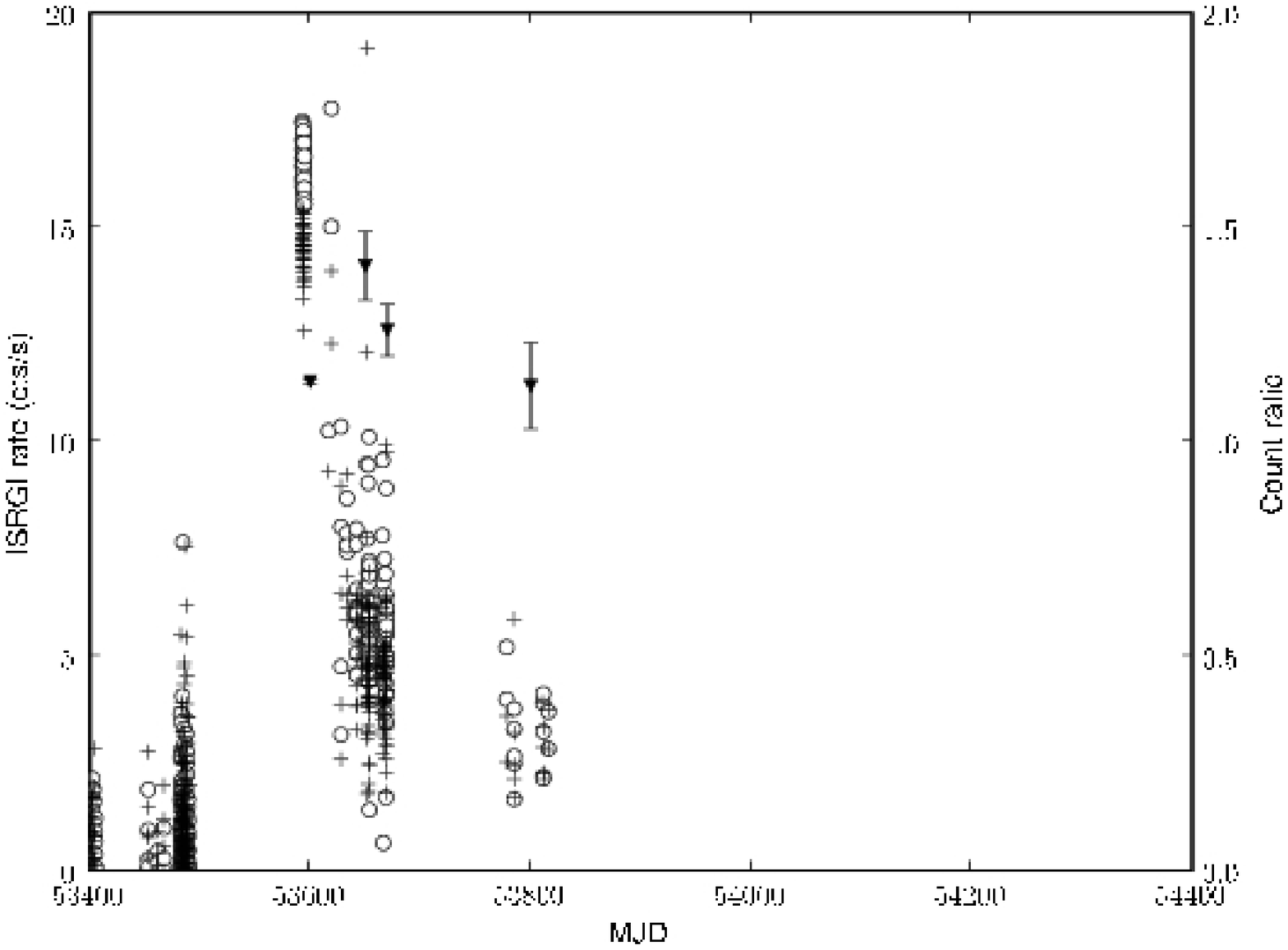}
\caption{(upper) RXTE/ASM long-term light curve \swift. The vertical dotted
  lines show the times of the observations by CB07, Miller et
  al. (2006) and our RXTE pointed observations in this work, from left
to right.
(lower) INTEGRAL monitoring average count-rates on the same
time-scale. Open circles are counts in the 22--40\,keV range, crosses
in the 40--80\,keV range (left scale), and triangles with errors are
average count ratios (22--40/40--80, right scale).}\label{ASM}
\end{figure*}

Figure \ref{ASM} shows the long-term luminosity trend of \swift\, as
seen in the X-ray and gamma-ray bands. 

It  appears that, after the initial
outburst and fade, the source has been steadily increasing in flux to
a peak at the time of our observation, of the order of the flux in the
tail of the initial outburst. Note that on the left-hand extreme of
Figure \ref{ASM}, one can see the zero level for the default
extraction for this source. The ASM count rate at the time of the June
2007 observations is clearly higher than this, near 
2\,cts\,s$^{-1}$. This corresponds to a flux of order $F\sim
1\times10^{-9}$\,erg\,s$^{-1}$cm$^{-2}$ in the 1--10\,keV range
(estimated  using WebPIMMS). Note further, that the INTEGRAL rate
ratios in the observations available were similar in each epoch, and
the light curve follows that of the ASM,
indicating that the high-energy portion of the emission changed little
in spectral slope/hardness. Unfortunately, observations closer in time
to our campaign are not yet publicly available.

In the optical, as detailed in Zurita et al. (2008), the counterpart
also settled to a steady optical magnitude, which has periodic and
aperiodic variability superposed.

\subsection{Orbital Modulation}\label{mcp}
\begin{figure*}
\begin{center}
\large{NOT}\\~\\
\includegraphics[width=0.45\hsize]{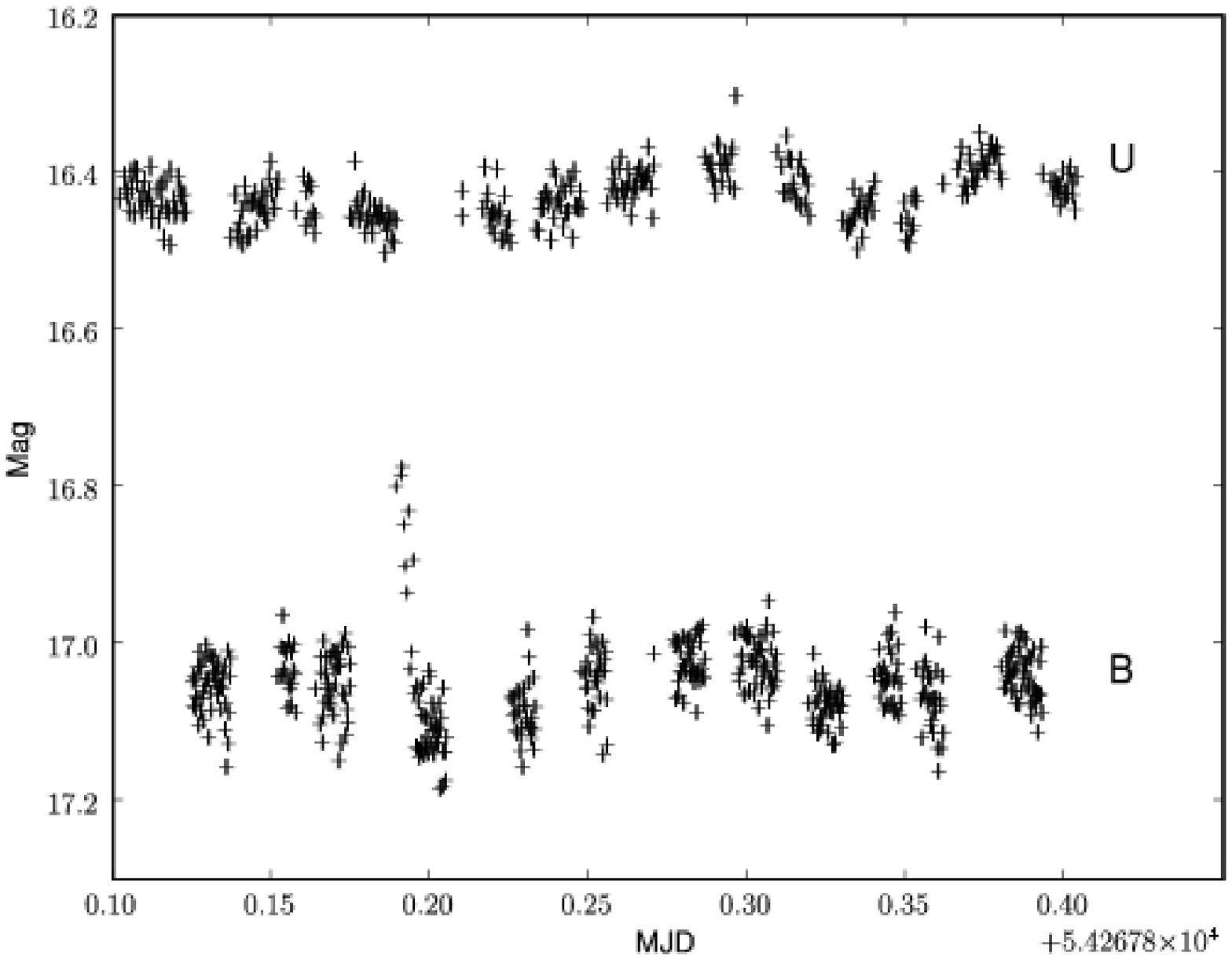}
\includegraphics[width=0.45\hsize]{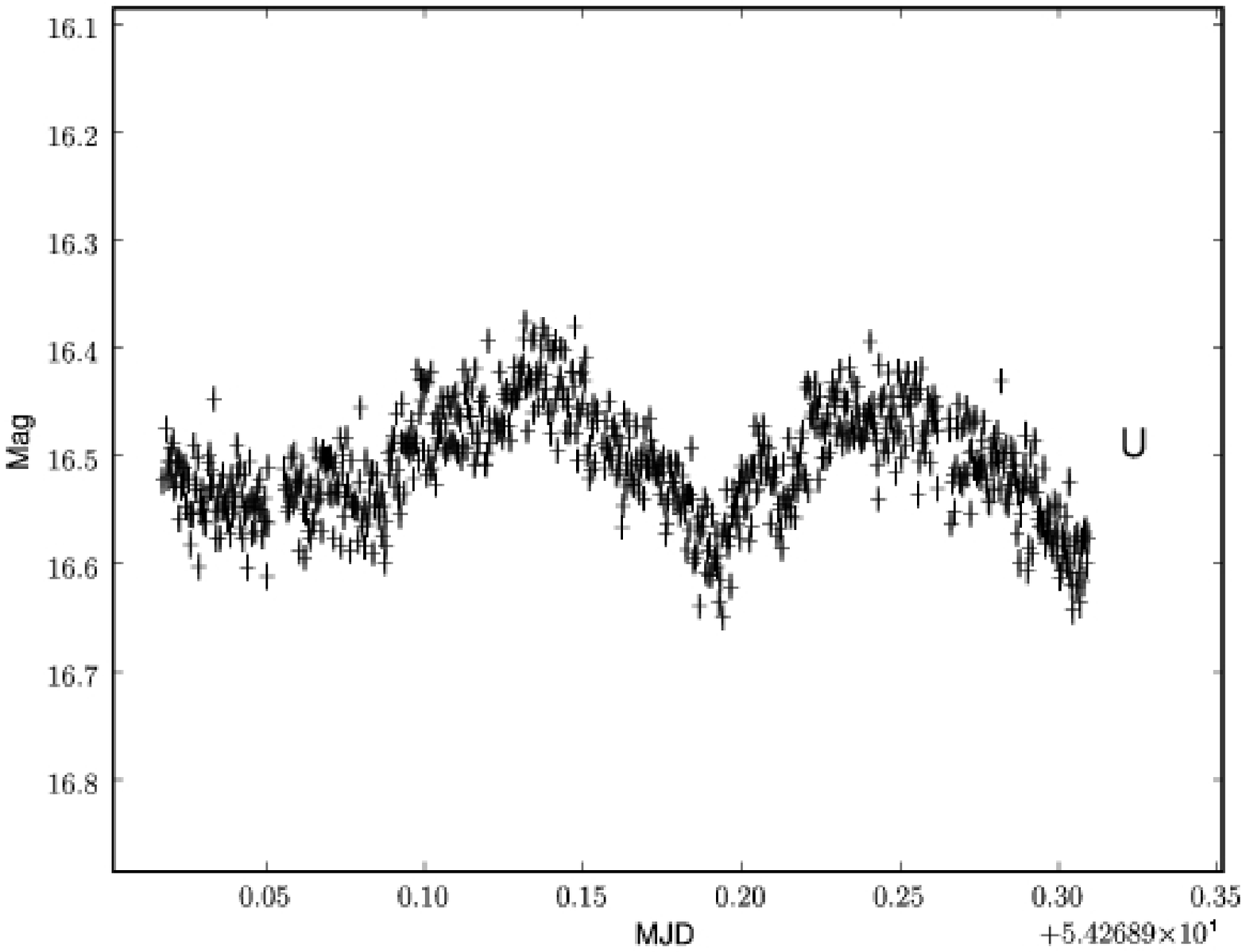}\\
\large{Mercator}\\~\\
\includegraphics[width=0.45\hsize]{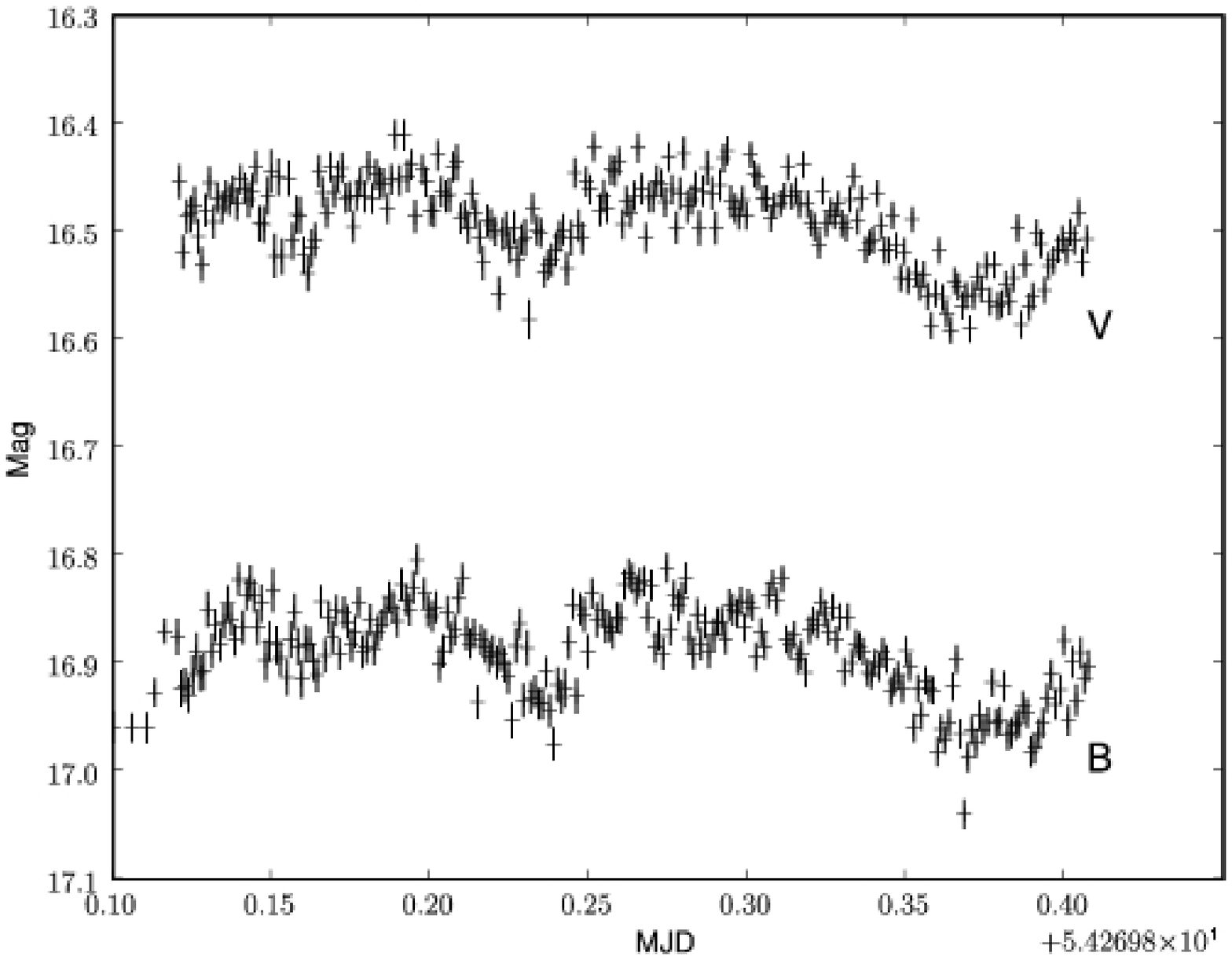}
\includegraphics[width=0.45\hsize]{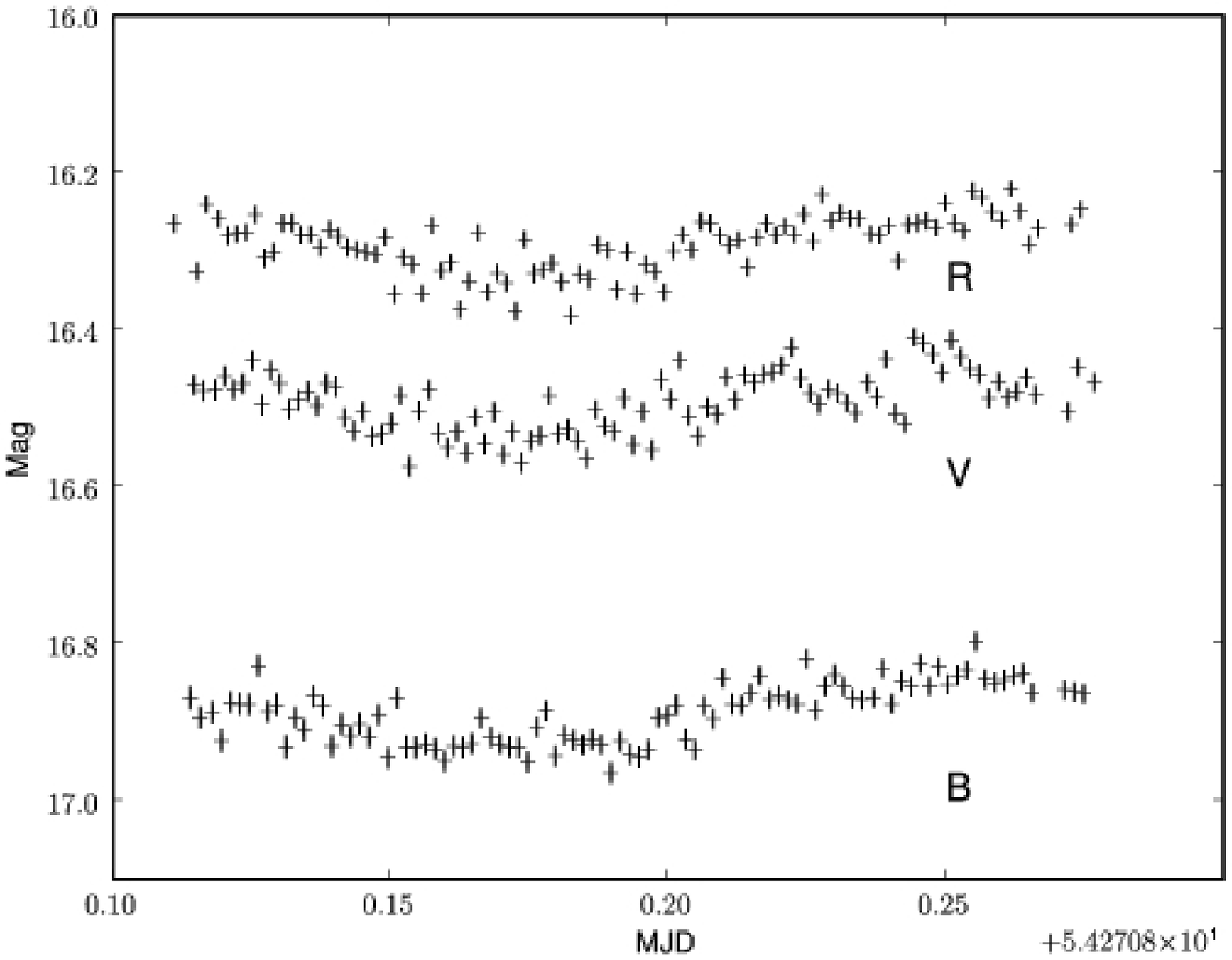}
\includegraphics[width=0.45\hsize]{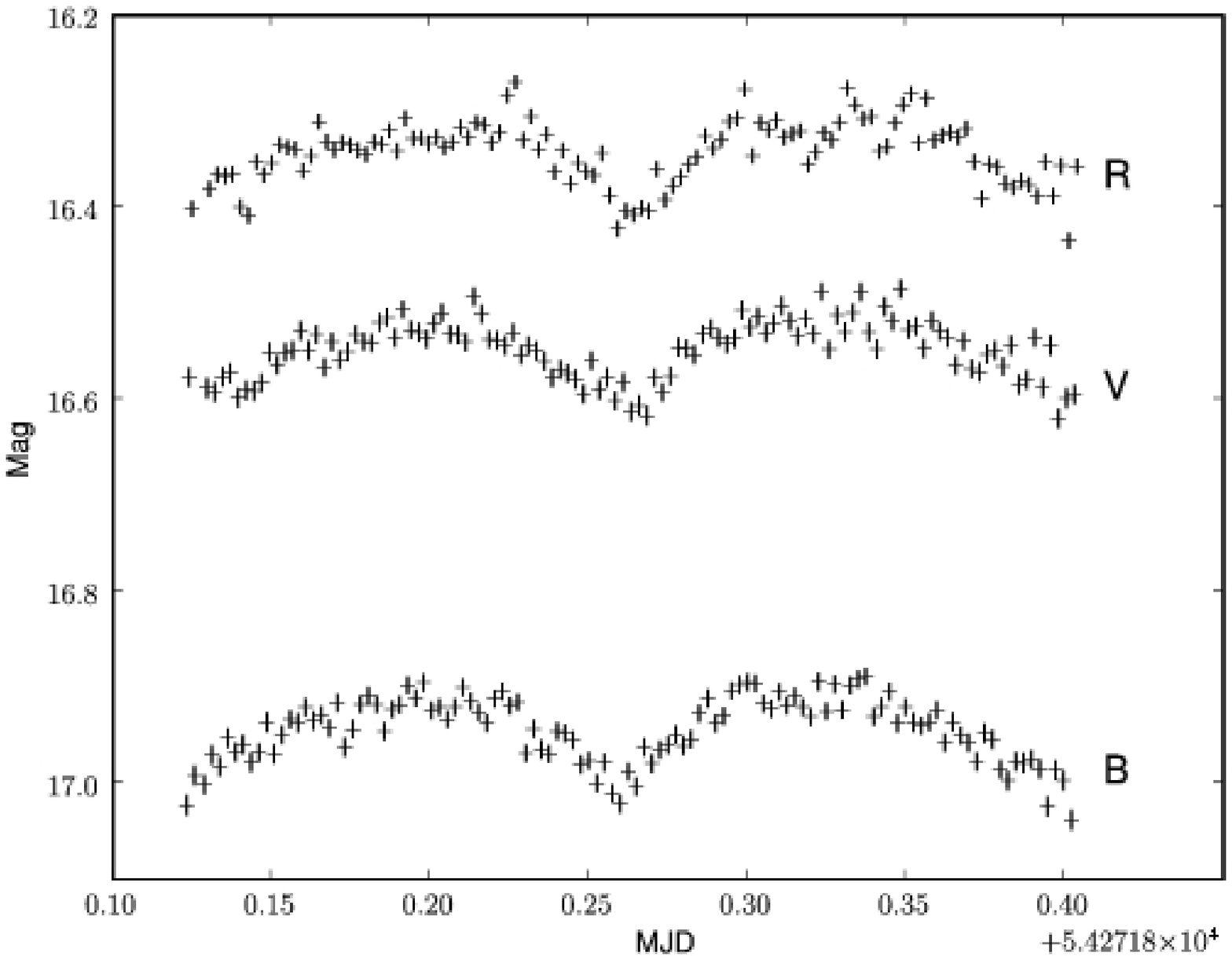}
 \caption{Optical light curves of \swift. The upper panels are from
   the NOT, the lower from the Mercator telescope. The five panels
   are subsequent nights 16--20 Jun 2007 (left to right, top to
   bottom). { Vertical lines are estimates of the statistical error on
   each point (but horizontal bars purely mark the point
   centre).} }\label{phot_lc} 
\end{center}
\end{figure*}

The light curves from our photometric observations with the NOT and
Mercator telescopes are shown in Figure \ref{phot_lc}.

In each night and each band ($UBVR$), the $\sim3.2$\,hr modulation
reported in the $R$ band by Zurita et al. (2008) is clearly apparent,
confirming their result. We are unable, however, to aproach the
accuracy of their period determination from our four nights'
data. Furthermore, simultaneous light curves in 
each filter show remarkably similar shapes. (Note that the
observations here were not strictly simultaneous: they were made
through each filter in turn) This variabilty has a shape typical of
superhump modulations (the conclusion reached by Zurita et al), but we
cannot exclude eclipsing (of the disc) or emission from the irradiated
side of the donor star from these data alone. Clearly, the colours vary
very little through the modulation period. The overall brightness does
vary during each night and from one night to the next. Table
\ref{photvalues} lists the mean magnitudes and colours from each of
the five nights. The variability within each night is smooth and at
the $\sim10$\% level.

\begin{table*}
\begin{center}
 \caption{Slow photometry of \swift.}\label{photvalues}
 \begin{tabular}{lcccc}
  \hline
Date (start)& $<U>$ & $<B>$ & $<V>$ & $<R>$\\
  \hline
2007-06-16 & 16.43(15) & 17.05(9) \\
2007-06-17 & 16.51(19) \\
2007-06-18 & & 16.89(15) & 16.49(14)\\
2007-06-19 & & 16.89(12) & 16.49(10) & 16.29(8)\\
2007-06-20 & & 16.94(13) & 16.55(11) & 16.34(11)\\
  \hline
 \end{tabular}\\
All magnitudes are in the Johnson-Cousins system. Numbers in
parentheses indicate the spread in values.
\end{center}
\end{table*}

Interestingly, our average magnitudes do not
match those of Zurita et al (2008), which were determined on 
2007 June 7, not far in time 
from our own observations. We find that the object is typically
brighter by $\sim0.1$\,mag in each of the bands $BVR$ in our data.

\subsubsection{Pulsed Fraction}

\begin{figure*}
\begin{center}
\includegraphics[width=0.7\hsize]{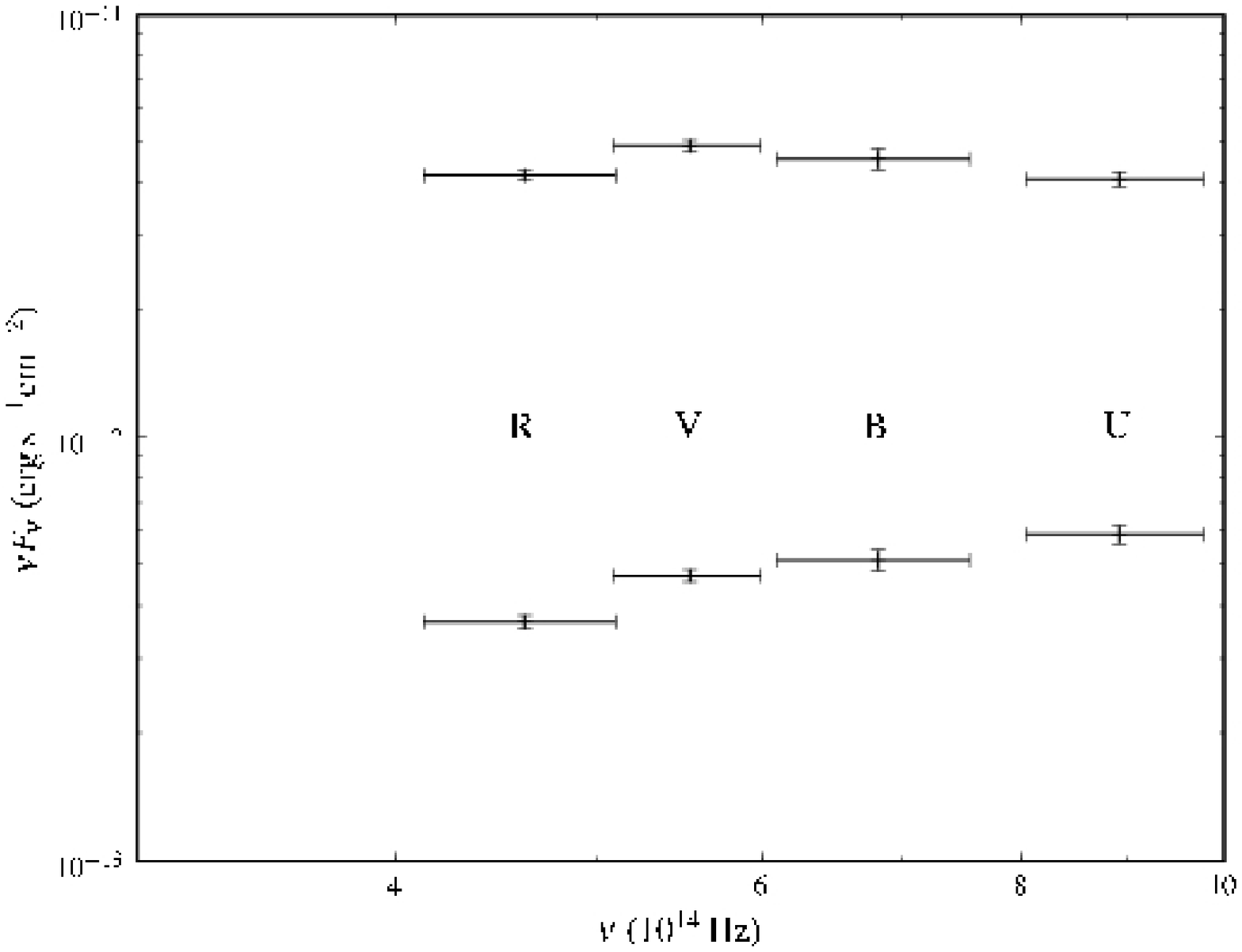}
\caption{Spectral energy distribution of \swift\ in the optical
  region. The upper points are the mean observed flux in the broad-band
  filters, with uncertainties derived from the scatter between nights. The
  lower points are the pulsed flux in the 3.2\,hr periodicity, with 
  uncertainties from the error on measuring the pulsed
  fraction. The points are as measured, uncorrected for
  reddening.}\label{pulse} 
\end{center}
\end{figure*}

The light curves presented also give information about the amount of
flux involved in the 3.2\,hr modulation. This can be measured directly
from the graphs from the peak-to-peak amplitude. In Figure \ref{pulse}
we plot both the total flux (as 
an average across the observations) and modulated flux in the optical
region. Here the uncertainties in the pulsed flux come from the
scatter in points on the light curves rather than variability between
nights - a caveat for comparing the points. Although there is
night-to-night variability, it is clear that the pulsed flux increases
with energy, whereas the total flux is flat or turns over. 
It appears that the pulsing component is hotter
than the DC component (or fast-varying component); { specifically, a
linear function fitted to the log-log points in Figure \ref{pulse}
yields slopes of -0.1(2) and 0.7(2) for the total and modulated
components respectively. } We can speculate that modulated emission
would dominate in the near-UV. Note that these 
points have not been de-reddened in order not to add additional
uncertainty to the plot. De-reddening the pulsed flux points by
$A_V\sim 1$ would yield a spectrum which is roughly consistent
with a Wien slope. The dependence of the pulsed fraction on wavelength
rules out that this component originates from X-ray reprocessing.

\subsection{Spectrum}
\subsubsection{X-ray}
\begin{figure*}
\includegraphics[width=0.6\hsize,angle=270]{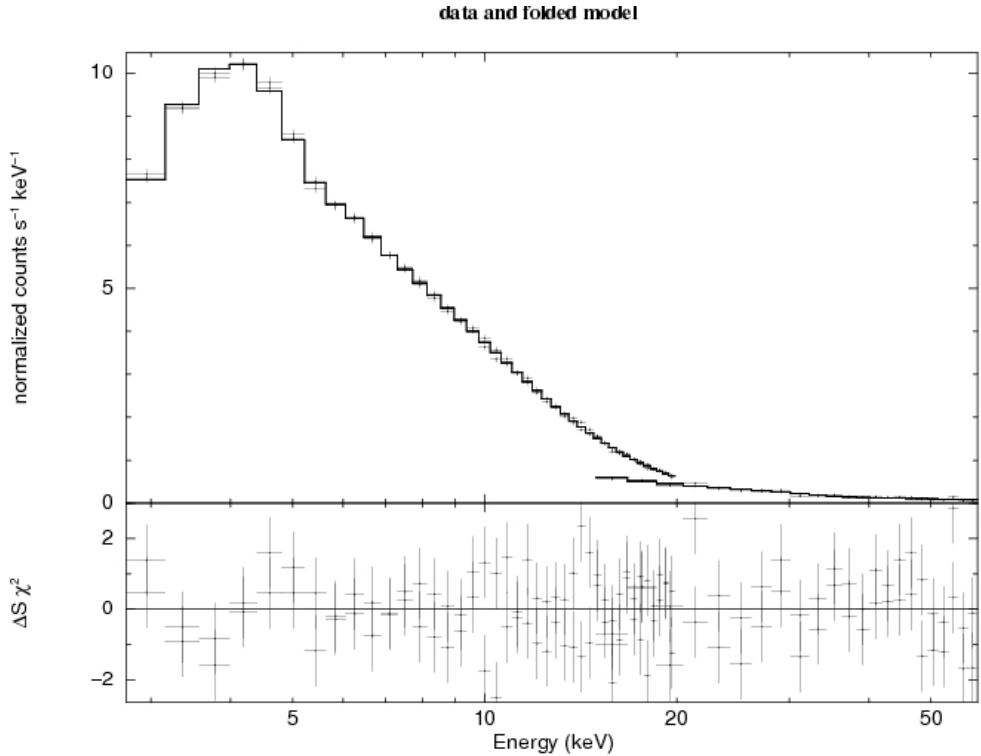}
 \caption{X-ray spectrum of \swift\ from RXTE for the two
   observations and two instruments: PCA (left) HEXTE (right). The solid line is
   the best-fit model in Table \ref{fits}.
}\label{Xspec} 
\end{figure*}

The spectrum was extracted from the valid photon events, and the
nominal background subtracted. We fit the resulting spectrum (using
{\tt Xspec})  in the range
2--50\,keV (including both PCA and HEXTE observations) to an absorbed
power law, and find a statistically 
satisfactory fit with $\Gamma=1.548\pm0.005$ ($N_\gamma \propto
E^{-\Gamma}$) and normalization 
$A=0.1057\pm0.0011$\,ph\,keV$^{-1}$cm$^{-2}$s$^{-1}$ -- see Figure
\ref{Xspec}. A thermal disc component at low energies and a Gaussian feature at
an energy $E\sim6.2$\,keV are consistent with the data and modestly
improve the $\chi^2$ statistic of the fit, each with a significance of
$\sim2\sigma$, after taking into account the uncertainty on the column
density for the former.  The broad line
might be consistent with a Fe K$_\alpha$ line with a rest-frame energy of
$E\approx6.4$\,keV. 

Specifically, using the hydrogen column found by CB07, we fitted the
X-ray spectrum with a model consisting of a power-law continuum as
above, summed with a disc thermal model ({\sc diskbb} in {\tt Xspec})
and Gaussian emission feature. The best-fit parameters we find are as
follows: disc temperature $kT = 0.19\pm0.03$\,keV,
normalization\footnote{defined in Xspec as 
  $\left[\frac{(R_{in}/\textrm{km})}{(D/10\,\textrm{kpc})}\right]^2$
  where $R_{in}$ is an 
  effective inner radius, from Kubota et
  al. (1998)} $N=4.8(6)\times10^4$, emission peak energy $E=6.15\pm0.17$\,keV,
normalization $A=2.63\pm1.3 \times
10^{-4}$\,ph\,cm$^{-2}$s$^{-1}$.  { The plotted points and fitting included only
photon-counting uncertainties and assume that the pipeline deals
correctly with the calibration. It should be noted, that the RXTE/PCA
calibration for the low energies of these putative detections is
somewhat uncertain, and further complicated by non-negligible
background contamination, which increases rapidly to lower energies
(Jahoda et al. 2006). The fit values and implied significance of the
components should therefore be read cautiously, particularly for the
thermal component.} We did, however, include the most recent
calibrations available, which fixed previously unknown problems with
the background estimation, such as correctly recording the time since
last passage through the South-Atlantic Anomaly \footnote{see {\tt 
http://www.universe.nasa.gov/xrays/programs/rxte/
pca/doc/bkg/bkg-2007-saa/}}.

We regard the detection of a disc component and of a broad
emission feature as suggestive, however the line would fit with a
gravitationally red-shifted, velocity-broadened Iron K line, as has
been seen for some other black hole accreting systems (Miller, 2007). For the
neutral Iron K line at $\approx 6.4$\,keV, the gravitational red-shift
would correspond to 3.6 Schwarzschild radii.

Ramadevi \& Seetha (2007) had found strong evidence for a very soft thermal
component in the X-ray spectrum of \swift\ during the brighter
emission near the outburst of 2005.  The
temperature of this component 
was $kT\approx0.4$\,keV, whereas Miller et al. (2006a) report a much
cooler thermal component of 0.2\,keV at a later time, further after the
outburst. The latter was based on XMM-Newton EPIC and
grating spectra, which are much more sensitive at low energies than
RXTE. Our marginal detection of a thermal component is
consistent with the later temperature. A comparison of preferred fit
parameters by different authors (at different times since the
outburst) are shown in Table \ref{fits}.

Note that Ramadevi \& Seetha (2007) reported finding an absorption
edge at $E\sim 7$\,keV.  In their analysis, this improved the
quality of the fit from $\chi^2_{red}\approx5$ to 1.3. This may
explain the departure from a power law in our spectrum. The shape
around this energy does not look like an absorption edge feature,
however. 

\begin{table*}
\begin{center}
 \caption{Summary of X-ray spectral fits and QPO detections by
   different authors.\label{fits}} 
 \begin{tabular}{lcccc}
  \hline
Parameter & Ramadevi \& Seetha & Miller et al. & Cadolle-Bel et al. &
This work\\
 & (2007)$^{1}$ & (2006)$^{2}$ & (2007) \\
  \hline
Satellite & RXTE & XMM & RXTE & RXTE \\
$T_{obs}$ & July-November & March & August & July\\
 & 2005 & 2006 & 2005 & 2007 \\
$\Gamma$ & $1.76\pm0.014$ & 1.66$\pm$0.01 & &1.548$\pm$0.005\\
$kT_{BB}$ (keV) & $0.38\pm0.07$ & 0.21$\pm$0.02 &
 & 0.19$\pm$0.03 \\
$R_{BB}$ $^3$ (km) & 100 & 2--6 & & 170\\
$kT_{seed}$ $^4$ & & 0.17$\pm$0.1 & 0.54$\pm0.06$ & \\
$\tau$ $^4$ & & 1.03$\pm$0.01 & 1.06$\pm$0.02 & \\
$\chi^2_{red}$ & 1.3 & 1.15 & 1.17 & 1.01 \\
$N_H$ (cm$^{-2}$) & $2.3\times10^{21}$ & $2.3\pm0.1\times10^{21}$ &
$2\times10^{21}$ &  $2.3\pm0.2\times10^{21}$\\
$f_{QPO}$  (Hz) & 0.891$\pm$0.008 & & 0.241$\pm$0.006 & See \S\ref{pds}\\
$\delta f_{QPO}$ (Hz) & $\sim$0.2 & & 0.03$^{+0.02}_{-0.01}$ &  \\
\%RMS & $\sim$23 & & 5.4$\pm1.8$ & \\
  \hline
 \end{tabular}\\
 Note that each fit is for a different
   combination of instrument, spectral range and fitted
   model, and that the hydrogen column is not necessarily variable in
   the fit. $\chi^2$ values are for the best fit in each case.\\
$^1$: Parameters at the start of the outburst.\\
$^2$: The power law plus black-body model, and the Comptomized
 model parameters both given.\\
$^3$: At 8\,kpc.\\
$^4$: Comptomized models, seed photon energy and corona optical depth.
\end{center}
\end{table*}

The total flux in the spectrum is
1.6$\pm0.1\times$10$^{-9}$\,erg\,s$^{-1}$cm$^{-2}$ in the 2--20\,keV
range. This is fainter than the flux reported by CB07,
which was measured in the tail of the initial outburst, and 
brighter than the flux reported by Miller et al. (2006), who performed
their observations some months after the outburst (the times of the
various follow-up observations are indicated in Figure \ref{ASM}, and
the fluxes measured match the ASM curve well).

\subsubsection{Optical}
\begin{figure*}
\begin{center}
\includegraphics[width=\hsize]{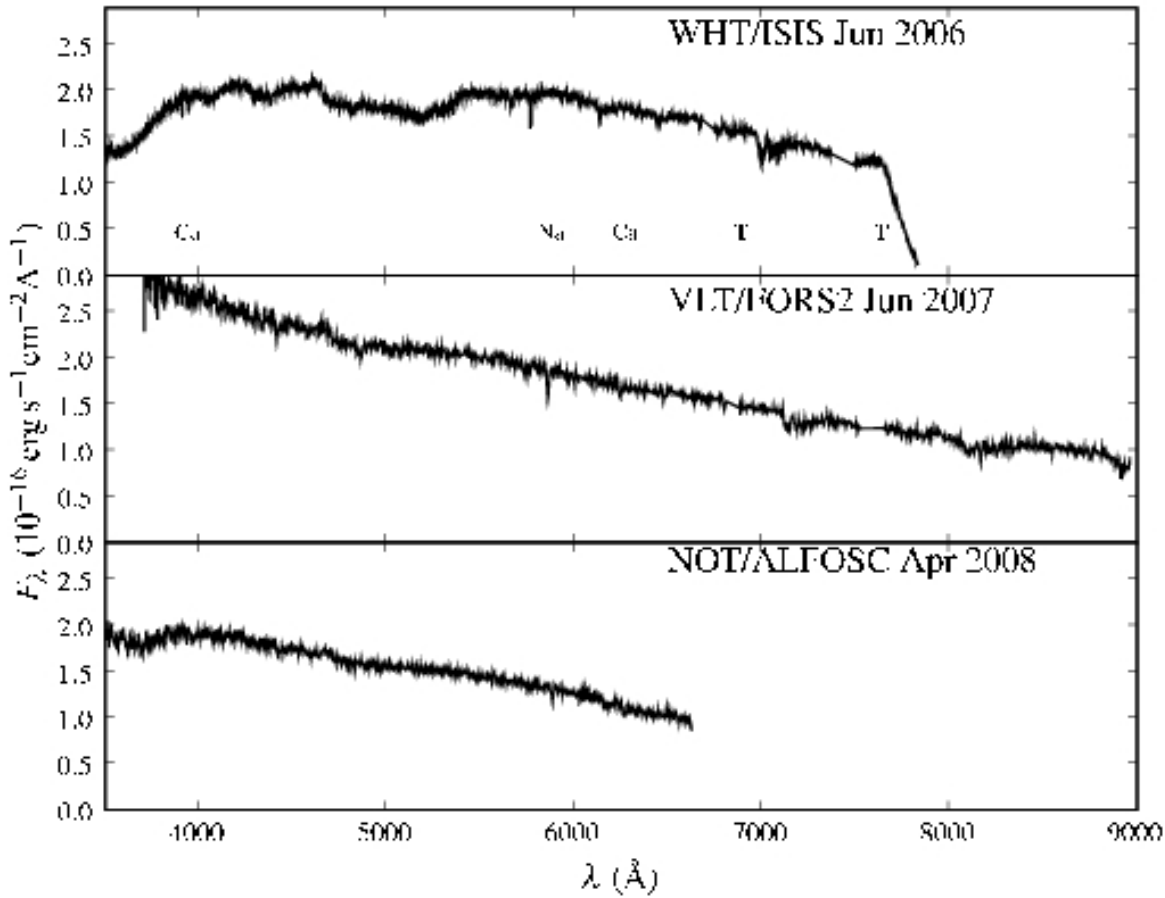}
 \caption{WHT/ISIS (upper),( VLT/FORS2 (middle) and NOT/ALFOSC
   (lower) optical spectra of \swift. A list of the equivalent widths
   of features in 
   these  spectra is given in Table \ref{lines}. The sharpest lines
   are marked, and atmospheric telluric absorption bands, which have
   been removed, are marked with ``T''.}\label{opspec}
\end{center}
\end{figure*}

The signal-to-noise ratio throughout the WHT spectrum (Figure \ref{opspec}) is
significantly better than for the later VLT/FORS2 spectrum, the latter
of which was taken in the same epoch as the rest of the
observations. Some of the features are apparent in both spectra, however, and
the equivalent widths for the most significant features is
given in Table \ref{lines}. Uncertainties are derived by measuring the
standard deviation of the flux in the continuum at either side of the
line in question. The continuum is assumed to be a straight line, and determined
from the data at either side of each spectral line. Assuming that the
continuum is a straight line and that the standard deviation is a
measure of the uncertainty on each point is clearly
not the case for such a undulating spectrum (in fact, it is an
over-estimate). Additional systematic
uncertainty due to this is not included in the errors presented in
Table \ref{lines}. 

{ The broad-band photometry measurements in Table \ref{photvalues} are
equivalent to average fluxes in their respective band-passes (e.g.,
the zero points of Bessel, 1995). These fluxes appear to be
some 20\% above the continuum level of our spectroscopy.} 
Note, however,  that the flux scale shown in Figure \ref{opspec} is affected by
slit losses and poor transparency during the observations. This
accounts for the apparent discrepancy between the typical flux shown
here and the broad-band magnitudes in Table \ref{photvalues}, which
are more reliable. { The photometry values show beyond doubt that the
spectrum shown has the correct shape (i.e., it is not affected by
broad-band mis-calibration of the continuum) and has reddened
signifiacntly since the optical spectrum of CB07.}

The Na-I interstellar absorption appears to be somewhat higher than
previously measured by CB07. Since this measurement is based on
relative fluxes (across the lines, compared to the continuum on either
side), and the intrument response functions do not change
significantly across the width of a narrow line in the centre of the
sensativity range, the statistical
uncertainties given should be a fair measure of the accuracy of the
EW measurements, affected only by the underlying  spectrum and not by
the instrument calibration, resolution etc.
Here we refer only to the WHT/ISIS
equivalent width, which should be more reliable. This implies a higher
reddening to the object and therefore correspondingly higher distance;
see below. { Since the interstellar extinction should not have changed,
we infer that the extinction internal to the binary varies. The
difference in the equivalent width between the WHT and VLT spectra is
$\Delta EW=0.4\pm0.2$ }

We can use the depths of the interstellar lines to estimate the total
extincting column to the source. Following the calculation in CB07,
$N_H = 0.25\times 5.8\times 10^{21}\times \textrm{EW}_{\textrm{Na\ I}} =
2.45\times 10^{21}$\,cm$^{-2}$. The uncertainty on this from the measurement of
the equivalent width alone is $0.13 \times10^{21}$\,cm$^{-2}$, but one
must keep in mind the uncertainty in determining the continuum and of
the conversion factors. We estimate that our column density is
accurate to 10\%.

From this we can in principle estimate distance to \swift\ by comparison with
field stars of known distance and extinction or H-II
clouds. Zurita et al's (2008) distance puts \swift\ well above the Galactic
plane, in the halo ($z>1$\,kpc). Unfortunately, at such a distance there are few
field stars, H{\sc II} regions or other fiducial distance indicators
to compare with. The bulk of the stars along the line of sight are
foreground Galactic Plane stars, so it is not possible to find the
main sequence or {\em red clump} at large distance in this
direction. The hydrogen column we estimate from absorption lines
above, larger than CB07's, is 
consistent with the total Galactic extinction (Still et al. 2006), so
the distance $d\sim$7\,kpc estimated by Zurita et al. from the lower column, 
is consistent with our analysis. { Note that there is no clear way to
determine how much of the extinction is internal to the binary
system. Hopefully, in a future quiescent state we would be able to
determine the donor type spectroscopically, and thus find the distance and
extinction independently. }

Beyond this, all features remain surprisingly weak, with a notable
lack of absorption, similar to the results of CB07, who see only a
faint hint of H and He absorption. Early in the outburst, there had been
significant H$_\alpha$ and He emission, initially double-peaked
(Torres et al. 2005), but there is very little evidence of
this in our data. For a line of similar width to the interstellar ones
seen, we place a 99\% upper limit on the equivalent width of
$EW<0.28$\AA, whereas Torres et al. found $EW(H_\alpha)\sim3$\AA\ (no
uncertainty given, except S/N$\sim$30). Finally,
the spectrum still appears blue (as seen 
also from the multi-colour photometry, Section \ref{mcp}), but less so
than at the peak of the outburst. 

The shape of the continuum in the VLT/FORS2 spectrum is qualitatively more
similar to the early spectrum in CB07 than to our intermediate
spectrum from WHT/ISIS. The blue part of the continuum appears to
have recovered somewhat, although the red has stayed rather similar
throughout. We note that, although much closer to a power-law shape,
the undulations or deviation froma  straight line by $\sim5$\%,
apparent in the FORS spectrum are real and not an artefact 
of the flux calibration (the calibration function was calculated for
each wavelength bin and smoothed, rather than attempting to fit with a
polynomial or other analytic function, which can easily introduce such
undulations). 

To check the evolution of the continuum shape, we performed further
optical spectroscopy with the ALFOSC intrument of the 2.5\,m Nordic
Optical Telescope (NOT), La Palma, in April 2008. The NOT/ALFOSC
spectrum, taken almost a year after our main observation campaign, is
also shown in Figure \ref{opspec}. Intriguingly, absorption 
lines are even less obvious in this spectrum than before, and we do
not list them; even the strongest Na-D line seems to have decreased in
strength. Using different telescopes, instruments and resolutions, it
is hard to say whether these changes are real. If so, they hint at a
fair amount of extincting material within the system, the amount of
which evolves with time. The undulations in the continuum (i.e.,
departure from a power law) are still
clearly present, however.

\begin{table}
\begin{center}
 \caption{Optical spectral lines in \swift. All equivalent widths are
   in absorption. DIL stands for diffuse
   interstellar line (of unknown, PAH or metallic origin). Limits are
   at 95\% significance.}
 \label{lines}
 \begin{tabular}{lccc}
  \hline
Element & $\lambda$ (\AA) & \multicolumn{2}{c}{Equivalent Width
  (\AA)}\\
 & & WHT/ISIS & VLT/FORS2 \\
  \hline
Ca & 3934 & 0.38$\pm$0.02 &<0.3\\
Ca & 3969 & 0.43$\pm$0.02 &<0.3\\
DIL & 4430 & 0.66$\pm$0.03 & 0.86$\pm$0.10\\
DIL &4885 & 1.00$\pm$0.06 & 1.15$\pm$0.11\\
DIL & 5780 & 1.09$\pm$0.05 &<0.6\\
DIL & 5797 & 0.222$\pm$0.007&<0.5\\
Na & 5892 & 1.69$\pm$0.09 & 2.1$\pm$0.2 \\
Ca & 6283 & 0.77$\pm$0.05&0.91$\pm$0.05\\
DIL &6614&1.20$\pm$0.11&<0.5\\
  \hline
 \end{tabular}
\end{center}
\end{table}

Slight undulations or humps are apparent in each of the spectra to a
greater or lesser degree. These are on the scale of a few percent in
flux and of order $\sim$500\AA\ broad in Figure \ref{opspec},
compared to a power law (straight line in these plots); such features
have now appeared in each of the four optical spectra since \swift's
discovery.  We must note that H and He emission lines {\em
  were} seen in the initial phases of \swift's outburst (Torres,
2005), so there is no lack of hydrogen or helium in the
system.

{The existance  of broad humps throughout the optical spectral range is
  suggestive of 
cyclotron emission, but offers  
little clear evidence; in theory they could be extremely broadened
lines or combinations of lines. The lack of sharp lines argues strongly that
the optical emission is not thermal emission from a dense disc
over-layed by less dense material, as this would result in absorption
and/or emission lines, depending on the dominant temperature. A pure
black-body or multi-colour dics would have smooth power laws either
side of a single peak. Likewise, synchrotron may well exist in the
spectrum, but is generally thought to have a $F_\nu\propto\nu^0$ flat
spectrum. 
Charateristic strong cyclotrons humps are not seen, but this
is not a similar situation to the more obvious cyclotron emission
in the case of Schwope et al. (2003): that is for a
polar system, where the emission region has a well-defined magnetic
field strength. In our case there could  be combined emission
from various active patches with various field strengths throughout
the disc, not necessarily resulting in a series of clear, well-defined
humps.

Thus although there is no clear evidence for cyclotron
emission, the spectra do not look like what might be expected from
other typical emission mechanisms, so cyclotron may be a successful
description. }

\subsubsection{SED}
\begin{figure*}
\begin{center}
\includegraphics[width=0.75\hsize]{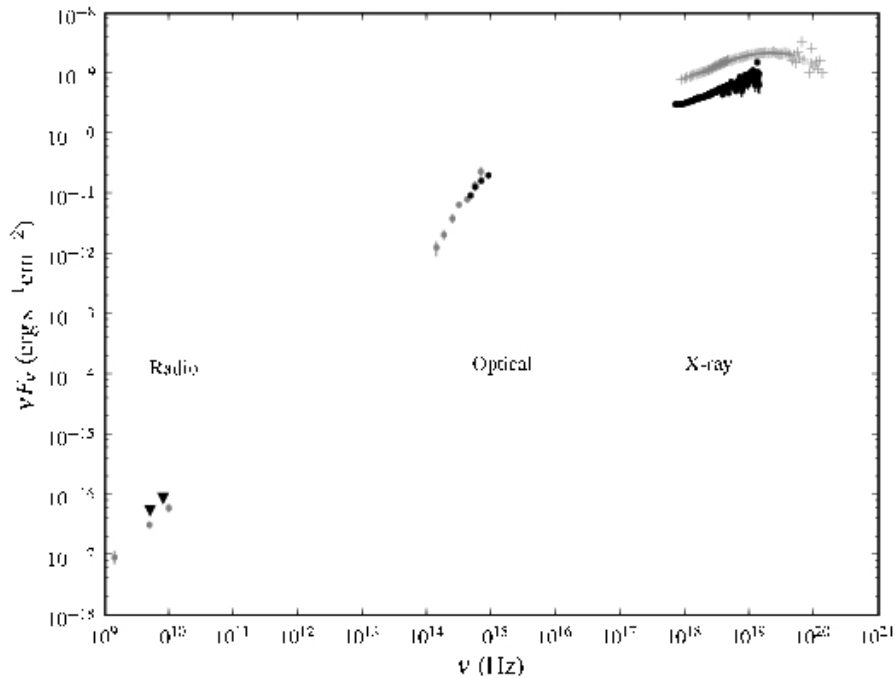}
\caption{Spectral energy distribution of \swift\ around the time of
  our observations (July 2007) from the radio to gamma rays. The radio
  points are upper limits. The grey points
are those of Cadolle-Bel at al. (2007; reproduced by kind permission)
the black points are from our
observations, black triangles are $3\sigma$ upper limits from Soleri
et al (2008). We have only included X-ray points with good
signal-to-noise, and de-reddened the optical by $A_V=1.05$ for a fair
comparison with Cadolle-Bel et al.'s points. }\label{SED}
\end{center}
\end{figure*}

In Figure  \ref{SED} we show the spectral energy distribution of
\swift\ in our observations, compared to the comprehensive
multi-wavelength campaign of CB07 over a year earlier. We have
included limits in the radio from Soleri et al (2008, taken two weeks
after our observations), and
de-reddened the optical with $A_V=1.05$, the value used by CB07 (so
that the comparison is fair). 

We find the overall 
luminosity has dropped by a factor $\sim$3, but that the X-ray spectrum has
qualitatively the same shape as before (a hard power law). The optical
is different: there appears to be a smooth break around the blue
such that the V-band flux is significantly fainter, but R and I are
consistent, and the slope is redder. One thing is clear: the
X-rays and optical require separate emission components, and if the
radio is at similar levels to before or lower by a similar factor as
the X-rays, then it too requires a separate emission
component: synchrotron emission from a jet cannot be dominant in the
optical region, if it has a typical flat $F_\nu$
spectrum. Intriguingly, the black body component suggested in the 
X-ray spectrum (and not present in CB07) may connect the softest
X-rays with the optical. This is merely speculation, however, as the
optical spectrum itself does not appear black-body-like.

{ Any radio emission, if synchrotron in origin (such as the flat $F_\nu$
spectrum in CB07) would be too weak to account for any significant
fraction of the optical emission.  Synchrotron does not, however, have
to be flat: emission from a single optically-thick clump rises as
$\nu^3$. Such a spectrum has not, to our knowledge, been seen thus far,
although de-convolving the various mechanism possible in the optical
is tricky.}

\subsection{Timing}
We produced light-curves for the X-ray observations of \swift. We used
only PCA rates, since the HEXTE
is far less sensitive. Note that there were three Proportional
Countrer Units (PCUs) active in the
second observation and only two in the first. Neither observation
shows any general or long time-scale 
trend in flux.

We also produced optical light curves, for further analysis.
For the optical light curve on the 17th, the
more stable of the two nights, rapid variability is seen, on the order
$\sim20$\,s, of a similar magnitude to the 3.2\,hr modulation
above. The typical uncertainty on each measurement is 0.03\,mag in
the $r'$ band, and 0.05\,mag in the $g'$ and 0.08\,mag in the $u'$ band
throughout the observation window.

\subsubsection{Power Density Spectra}\label{pds}

\begin{figure*}
\includegraphics[width=0.6\hsize]{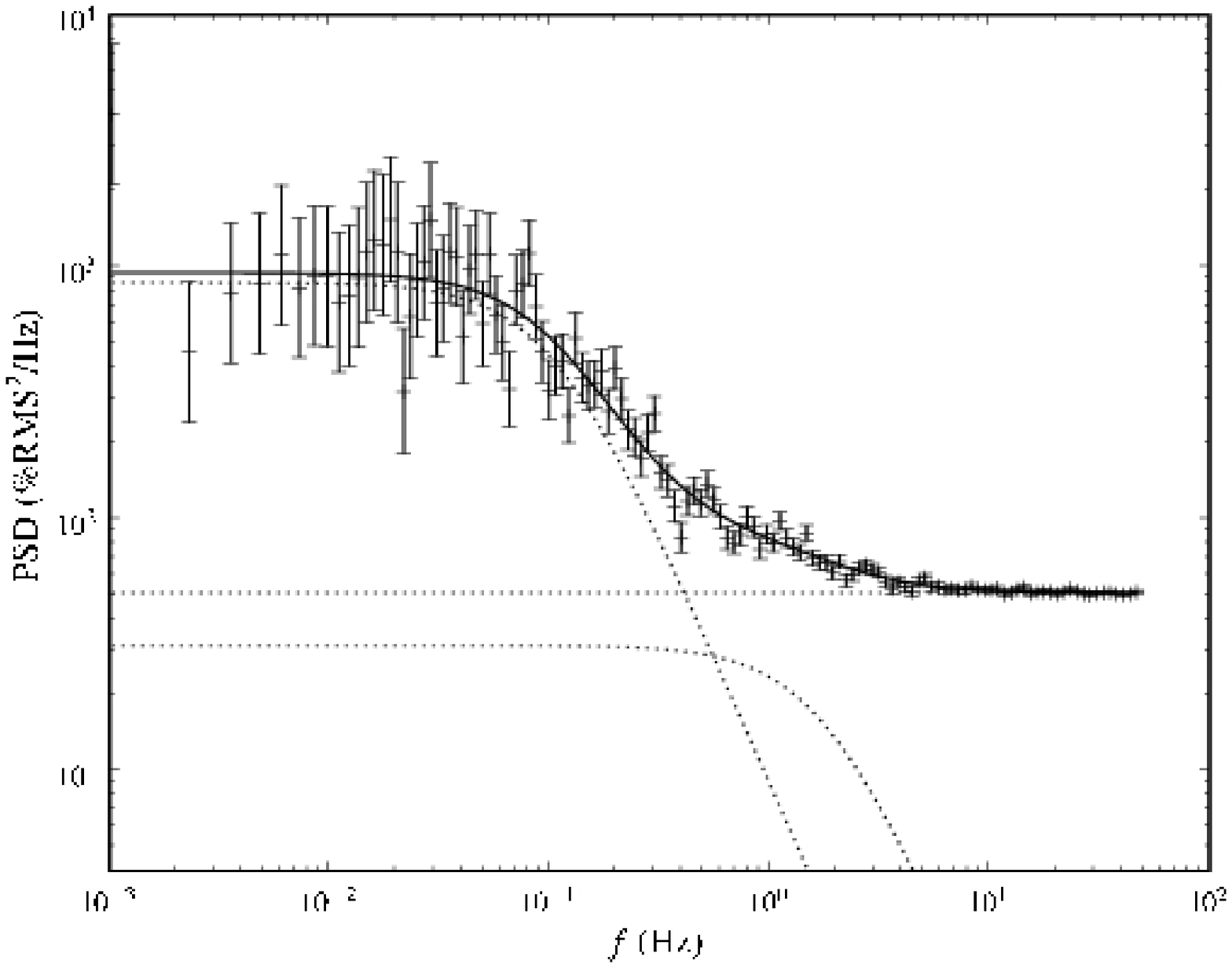}
\includegraphics[width=0.6\hsize]{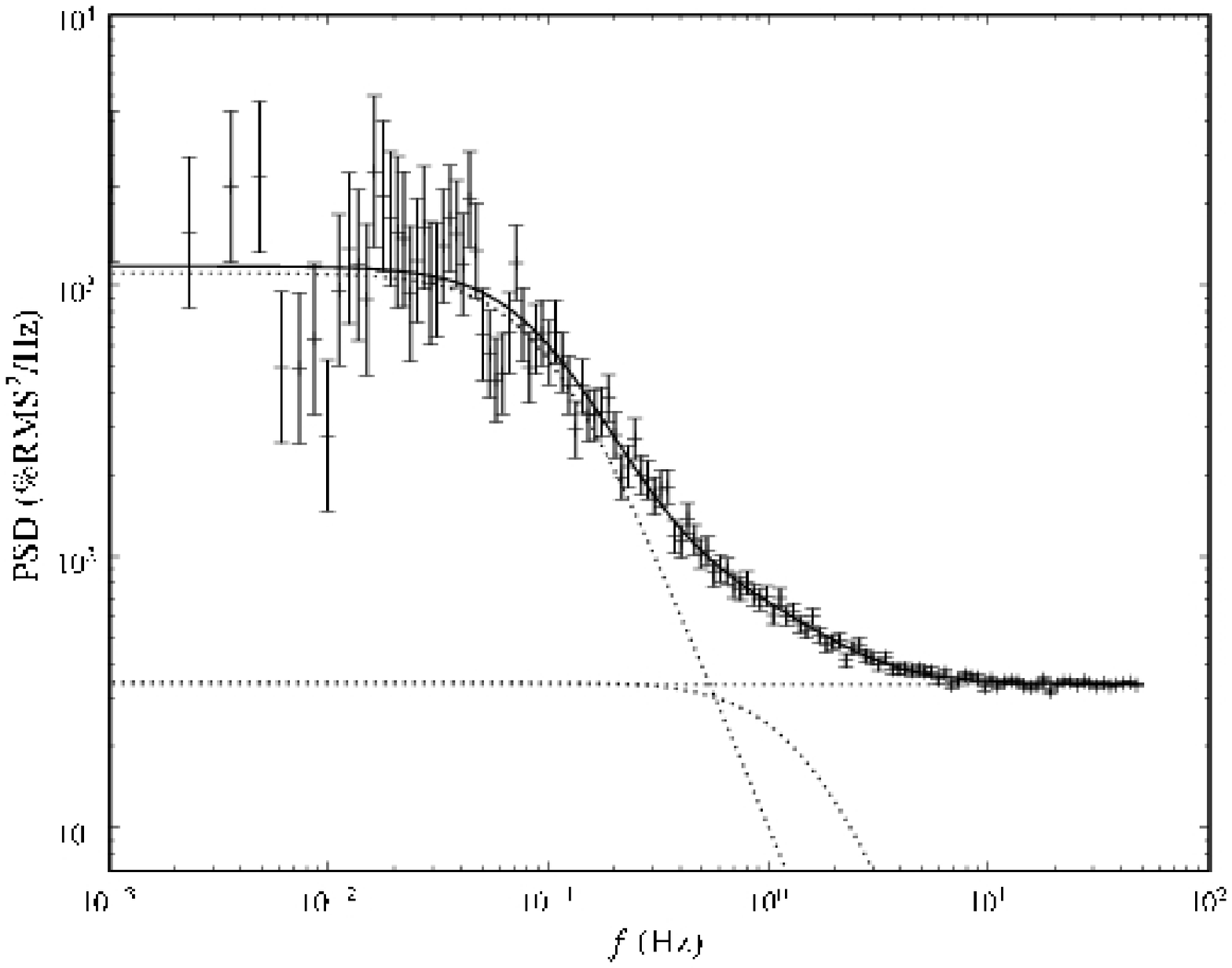}
 \caption{Power spectra of the
   RXTE/PCA observations of \swift, 2007 June 11 (top) and 13 (bottom).  The
   power spectra have been rebinned and the best fit over plotted
   (solid line) and its components (dotted lines, see Table
   \ref{pgramfits}).}\label{pgram1}\label{pgram} 
\end{figure*}

\begin{figure*}
\includegraphics[width=0.60\hsize]{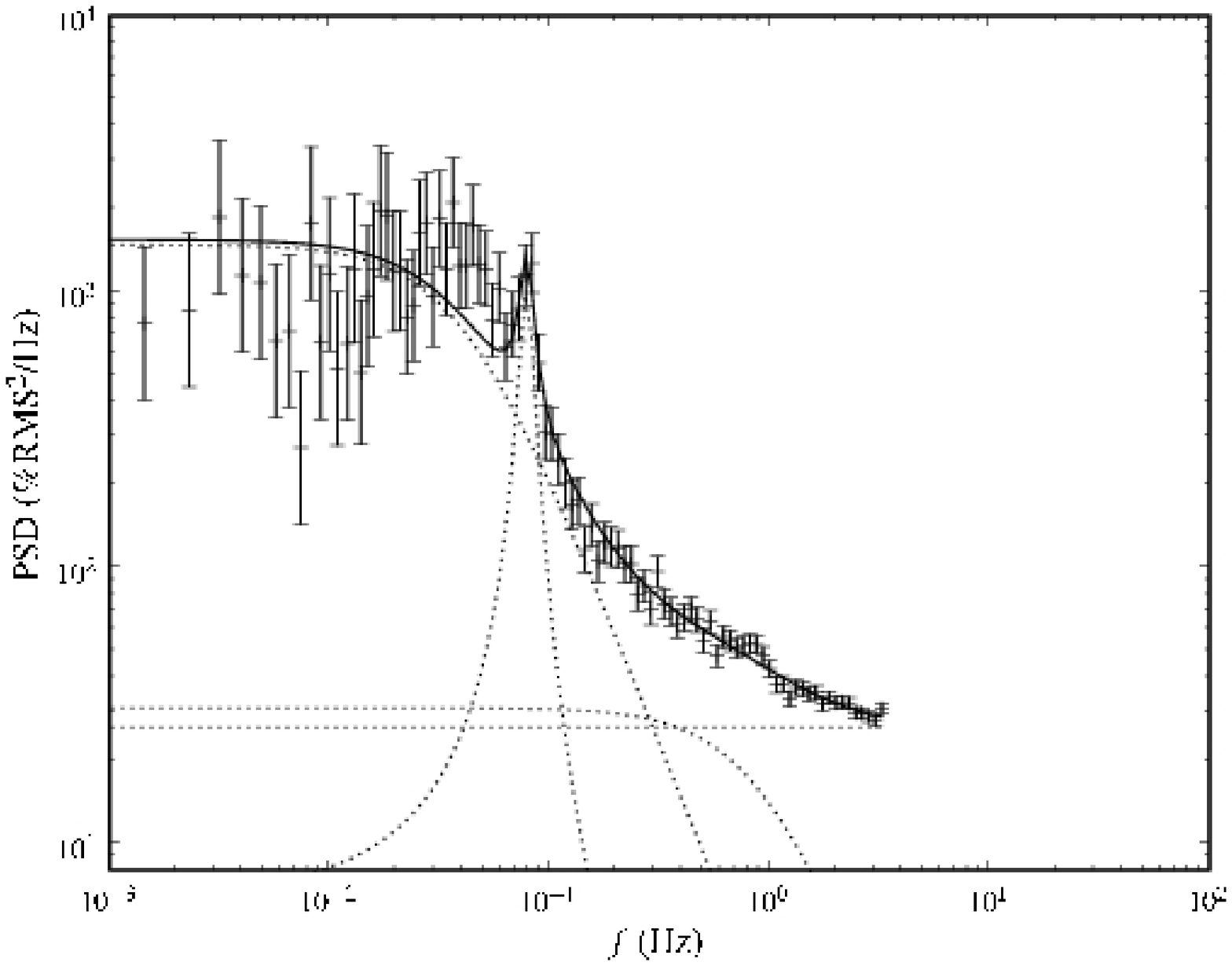}
\includegraphics[width=0.60\hsize]{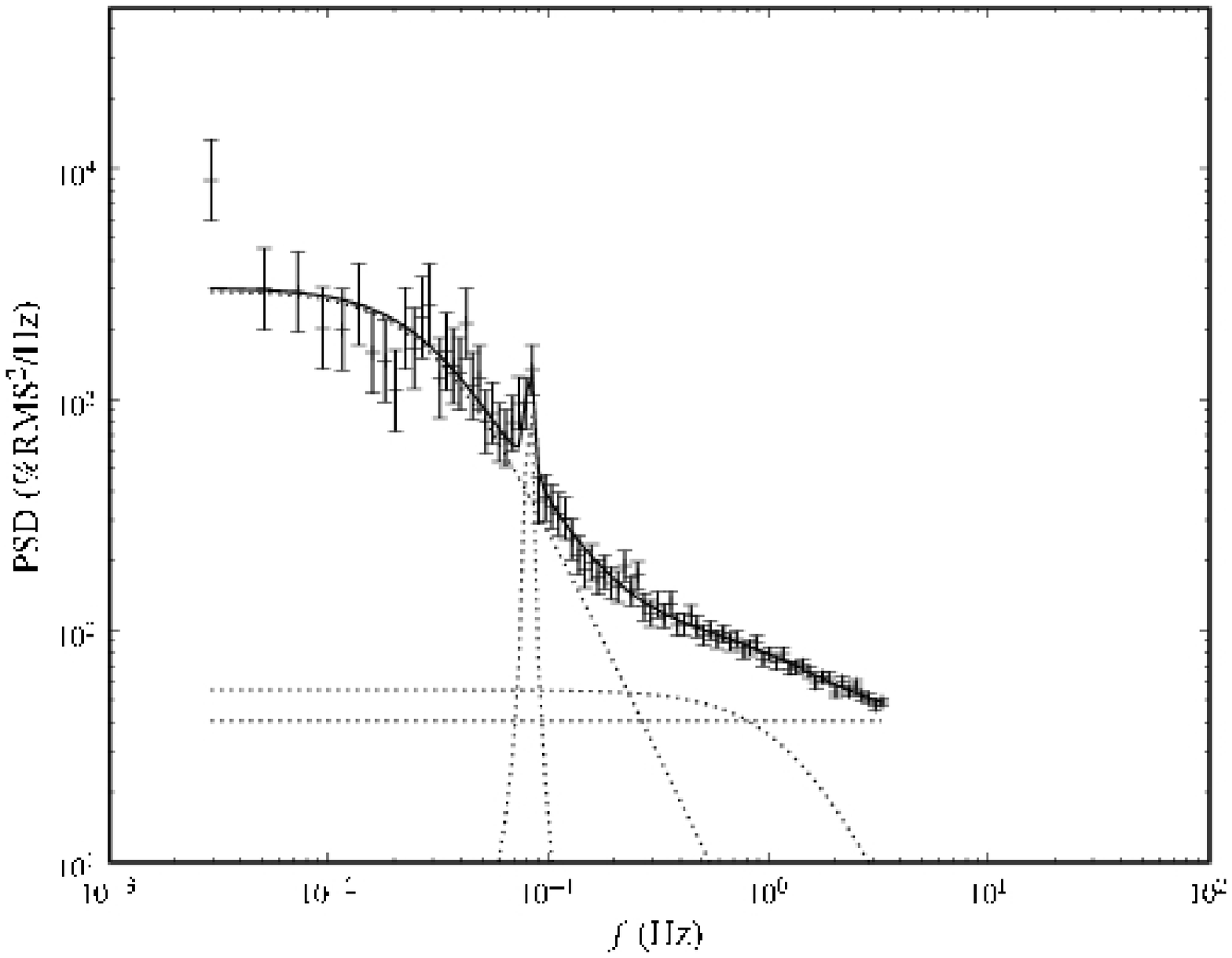}
\caption{Power spectra of the ULTRACAM
   observations of \swift from 2007 August 13, r' (top) and g'
   (bottom), displayed as in Figure \ref{pgram1}}\label{pgram2}  
\end{figure*}

\begin{figure*}
\includegraphics[width=0.60\hsize]{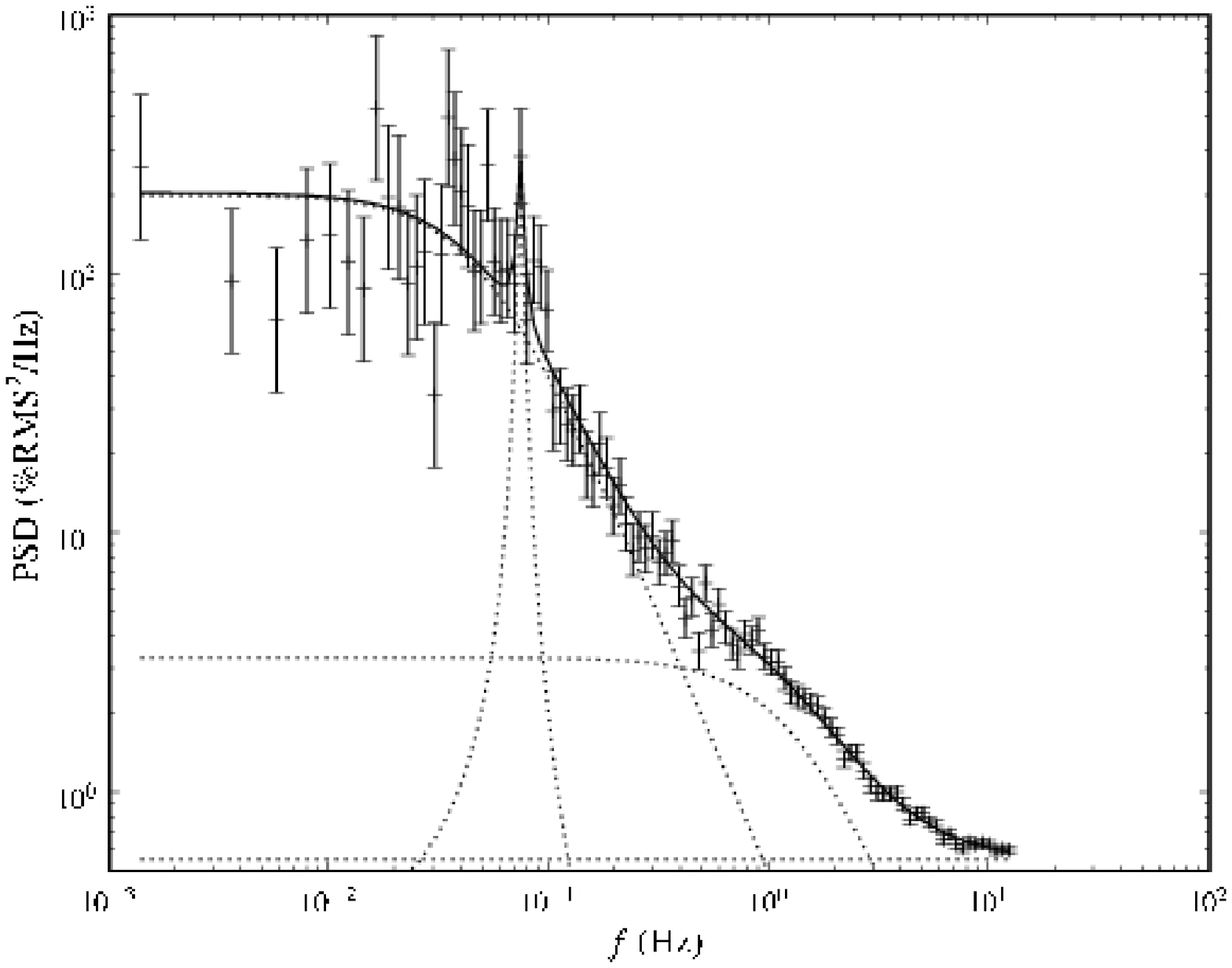}
\includegraphics[width=0.60\hsize]{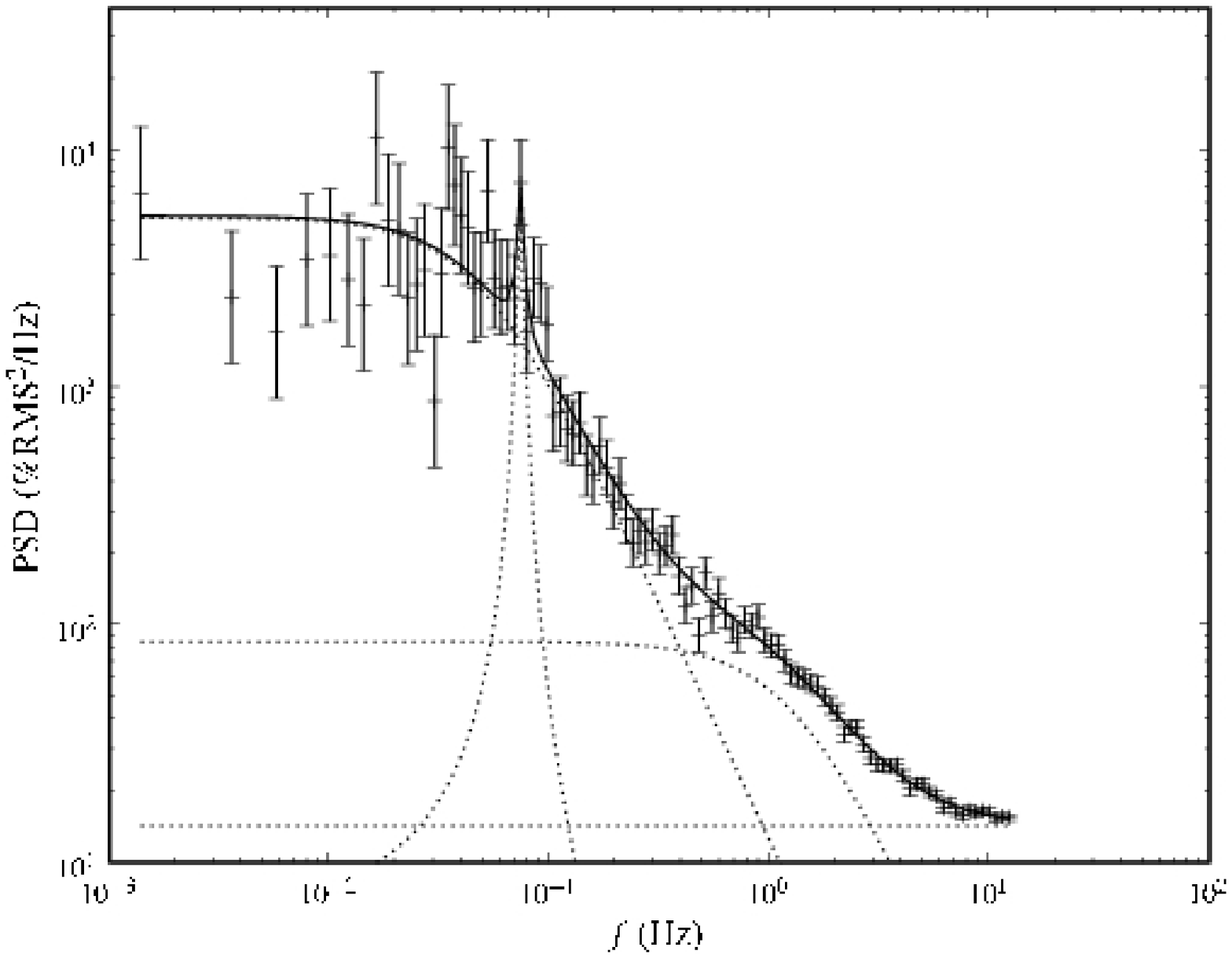}
\caption{Power spectra of the ULTRACAM
   observations of \swift from 2007 August 18, r' (top) and g'
   (bottom), displayed as in Figure \ref{pgram1}}\label{pgram3}  
\end{figure*}

\begin{figure*}
\includegraphics[width=0.60\hsize]{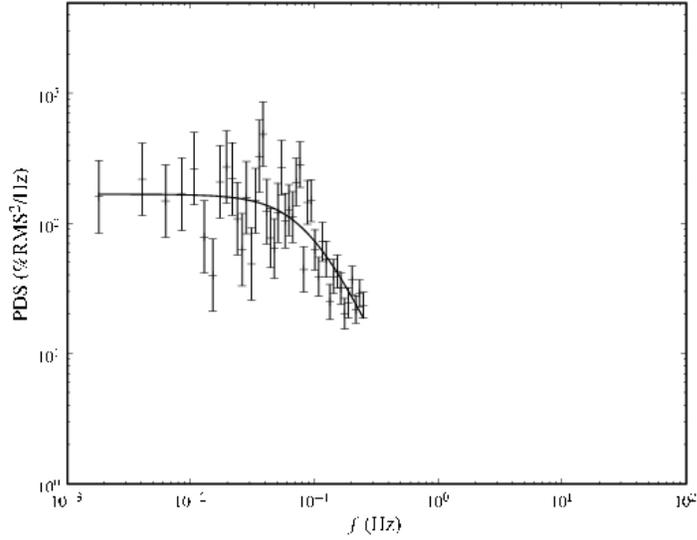}
\caption{Power spectrum of the ULTRACAM
   observations of \swift from 2007 August 18, u',
   displayed as in Figure \ref{pgram1} with one fitted
   Lorentzian.}\label{pgram4}   
\end{figure*}

A comparatively high fraction of flux is involved in short-term
($T<1$\,min) variability in both X-ray and optical -- 
$>$10\%, similar or larger than the orbital-like modulation above.
Figures \ref{pgram} to \ref{pgram4} show the power spectra for all the
light curves. Note that, for the optical, the
Lomb-Scargle method is required, because the sampling is not strictly
regular, and some points have null values. A Fourier analysis of these
data do give indistinguishable results, if one assumes regular
sampling. White noise can be calculated by assuming a standard
deviation equal to the Poisson noise
($\sqrt(N)$) on each sample, or obtained by fitting a function to the
power spectra which includes a constant term.
Fitting the power spectra and calculating from the known
count rates give very similar values for the white noise.

\begin{table*}
\begin{center}
 \caption{Power spectrum best-fit parameters. The power law (PL) exponent
 refers only to the high-frequency ($>$0.1\,Hz) part of the
 power spectra, whereas  the rest of the parameters are for double
 zero-centred Lorentzians  (see text).} \label{pgramfits}
 \begin{tabular}{lccccccc}
  \hline
Parameter & \multicolumn{2}{c}{RXTE/PCA} &
\multicolumn{4}{c}{ULTRACAM}\\
 & 2007 June 11& 2007 June 13& \multicolumn{2}{c}{2007 June 13} &
\multicolumn{3}{c}{2007~~ June~~ 18} \\
 & & & $r'$ & $g'$ & $r'$ & $g'$ & $u'$~$^1$\\
  \hline
PL exponent & 1.20 & 1.18 & 0.95 & 0.90 & 0.92 & 0.89 & 1.7\\
  \hline
Width 1 (Hz) & 0.102(9) & 0.095(8) & 0.040(4) & 0.033(3) &0.049(6) & 0.051(5) &
0.087(9)\\
\%RMS & 12.0(7)& 13.1(7) & 9.70(8) & 37.7(3) & 4.00(5) & 15.8(2) &
4.8(3) \\
Width 2 (Hz) & 1.7(2) & 1.5(2) & 0.9(2) & 1.5(2) & 1.3(2) & 1.3(3) & \ldots\\
\%RMS & 9.3(8) & 9.1(7) & 6.15(15) & 32.2(5) & 2.55(5) &10.7(3)& \ldots\\
QPO? (Hz) &\ldots &\ldots  & 0.0789(15) & 0.0791(16) & 0.079(2) & 0.074(12) &
0.08(3) \\
\%RMS & $<$3 & $<$3 & 4.4(10) & 11(2)& 1.5(6) & 7(2) & $<$10 \\
$\chi^2/dof$ & 145/123 & 116/123 & 136/85 & 101/85 & 92/95 & 94/95 & 55/42\\
  \hline 
 \end{tabular}~\\
Numbers in perentheses are 1-sigma uncertainties in the last digit, 
95\% confidence for limits.\\
$^1$~Due to the smaller number of points available, the fitting
function only included one Lorentzian. \\
\end{center}
\end{table*}

{ Qualitatively, one can see a number of features common to the
power spectra: a power law-like decay at high frequencies and a break
around 0.07\,Hz 
becoming flat in the 0.01\,Hz range. Some possible QPOs
can be seen at around 0.08\,Hz in the optical power spectra but not in
the X-ray ones, but with significance only at the
2--5$\sigma$ level. The optical power spectra are all similar to
one-another in this respect also. }

For the $u'$ band, we obtained a light curve by co-adding
groups of 50 images, and analyzing these. The temporal resolution was
thus degraded (to about 2\,s), with still significant scatter in the
measurements.  This accounts for the different appearence of the $u'$
band power spectrum in Figure \ref{pgram4} (the $u'$-band power spectrum
is the one with the least points). 
Many of the same features can be seen in the power spectrum: red noise
power law, break, excess in the 0.01--0.1\,Hz region and a flat power
spectrum at low frequencies. Even a QPO is possibly seen, but not
statistically significant.

By visual inspection, we find that the power spectrum continua are each
well-fitted by zero-centred Lorentzian functions (as used by CB07, for
example). We have, therefore, fitted each power spectrum accordingly, and
the result of these fits are shown in Table \ref{pgramfits}. Also
shown are what one would find assuming a power-law noise function for
the high-frequency ($>$0.1\,Hz) part of each power spectrum. These numbers are
consistent with {\em flickering} (superposition of discreet,
stochastic flares of various heights and durations), for which one
expects a power-law exponent of 1 (Bruch, 1992). The flattening of the
power spectrum towards lower fequencies does not imply that this
considerable power cannot be produced by micro-flares, but implies
that the micro-flares cannot have arbritarily long durations.

{ Uttley \& McHardy (2001) give arguments against a Lorentzian and/or
power-law power spectrum necessarily being produced by
flickering; specifically, if the average flux and variance of sections
of the light-curve are linearly correlated, without the line passing
through the origin. This is indeed to case here, and implies that
although flickers can occur on a wide range of time-scales as measured
by the power spectra, there is likely an extra component of constant or
very small RMS, perhaps the process which feeds the larger amplitude
variability (e.g., Merloni \& Fabian, 2001).

Although the power spectra all look rather similar, the numbers
presented in Table \ref{pgramfits} are not. If the ideas of Uttley \&
McHardy are correct, then there seems to be a variable competition
between the variable component and the low RMS/constant component
which depents both on time and spectral range. For the simultanous r'
and g' numbers, the variability is consistently higher for the shorter
wavelength, yet we see generally more variability in the optical than
in the X-rays (where, as already stated, we do not find significant
QPOs). Any micro-flares must thus be rather blue, but damped in their
effect on the X-ray emission.}

\subsubsection{Auto-correlation functions}

\begin{figure*}\begin{center}
\includegraphics[width=0.45\hsize]{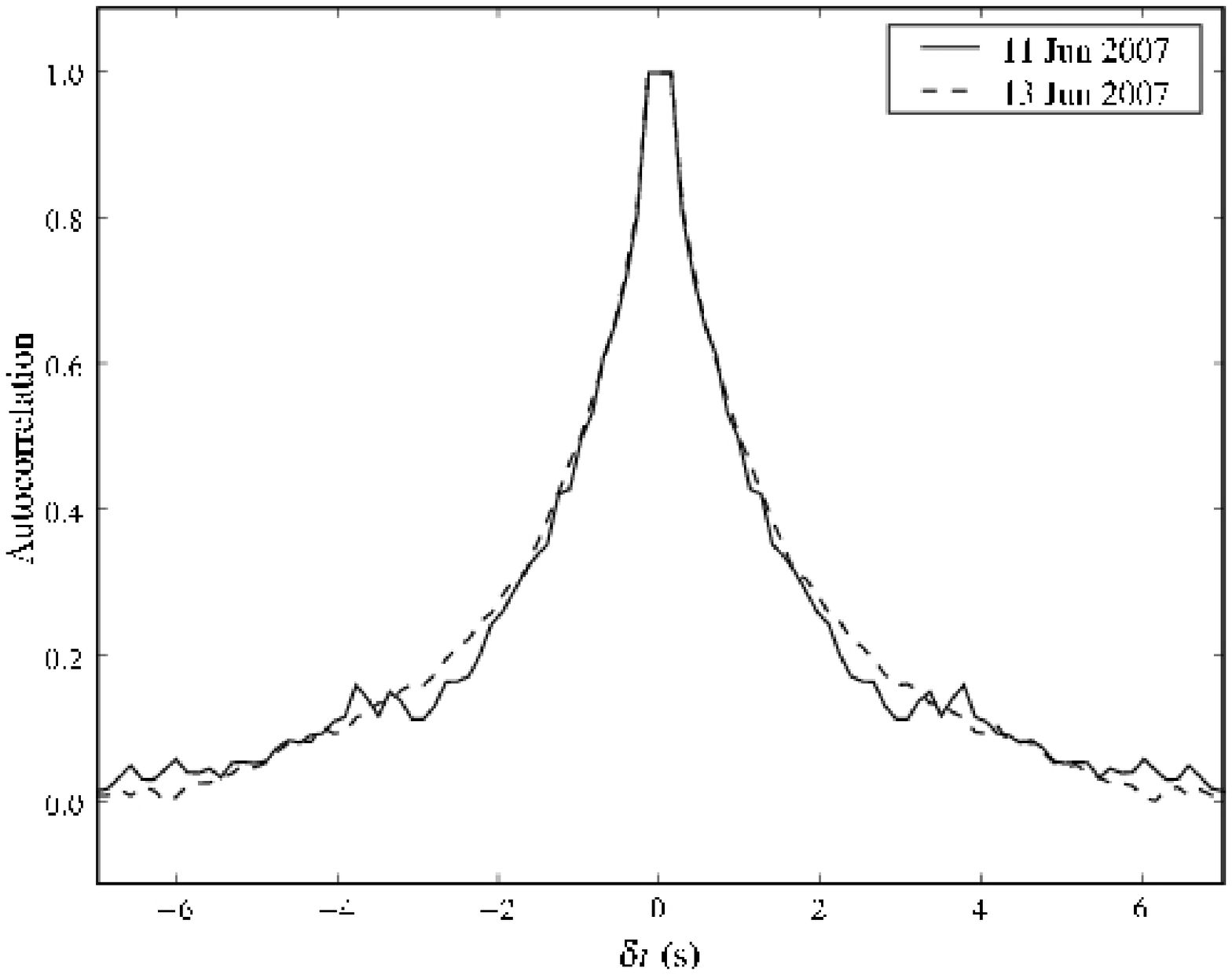}~\\
\includegraphics[width=0.45\hsize]{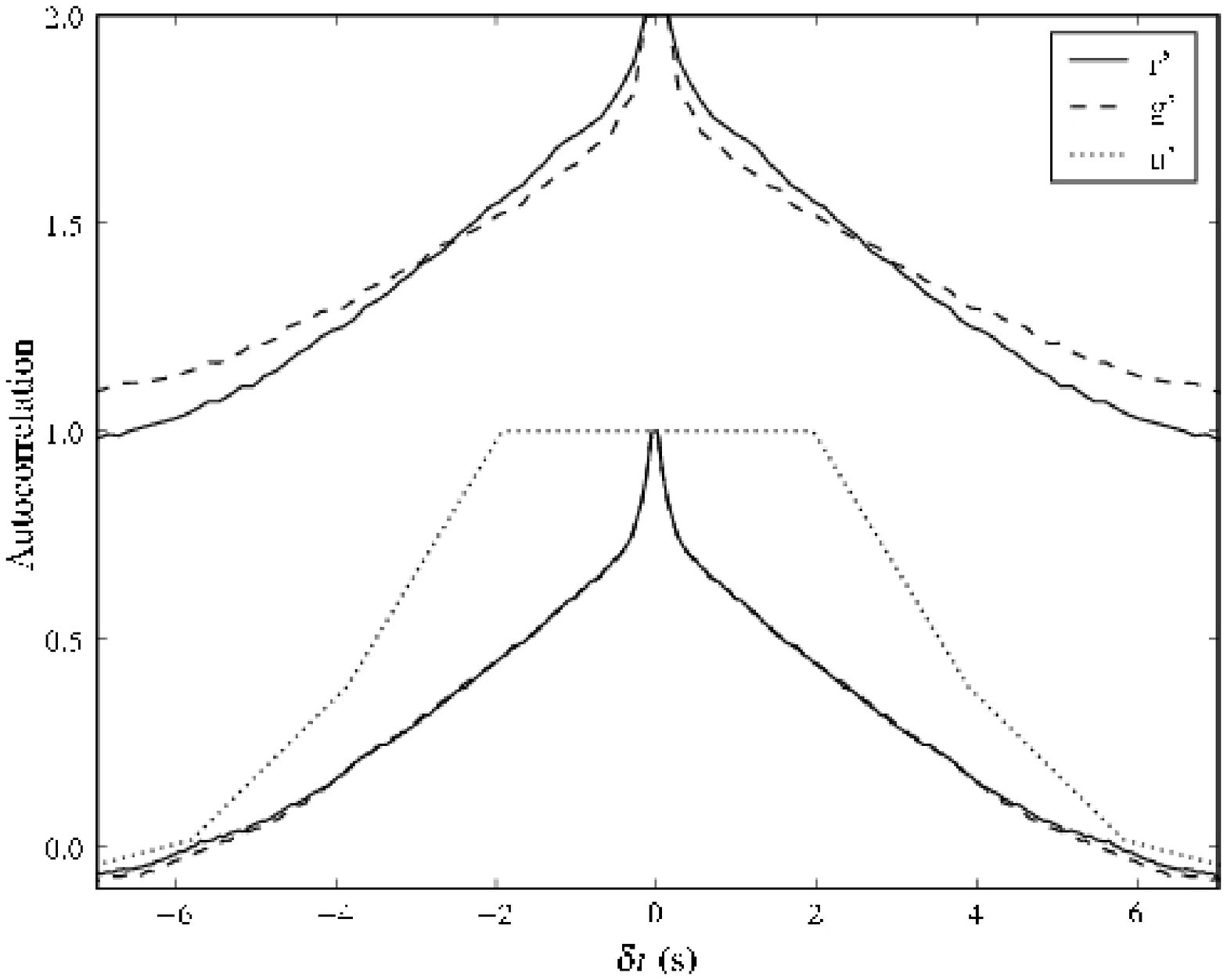}
\caption{Auto-correlation functions for the different
  light curves. The top panel shows the two RXTE observations, the
  bottom panel the optical for the two nights, with the 
  2007 June 13 data shifted vertically by $+$1.  Each
  function is normalized to 1 at the peak, and the central point at
  $\delta t=0$ has been omitted, as it is strongly affected by white
  noise. The $u'$ curve 
  represents a light curve of much lower time-resolution (by a factor
  of 50), and its apparently enhanced width is a result of
  this.}\label{autocorr}
\end{center}
\end{figure*}

Figure \ref{autocorr} shows the auto-correlation functions for the
fast timing observations of \swift. This is a measure of how well each
light curve is correlated with itself as a function of time
difference, and therefore of the timescales dominant in the
variability. In this sense, it complements the power spectrum, viewing
the time-series from the point of view of individual events rather
than coherent periodic signals. All the curves are well 
approximated by Lorentzian functions. The X-ray autocorrelation
functions are very 
narrow (FWHM$\sim 2$\,s), implying that the variability seen at
longer time-scales in the power spectra above are not very
coherent. The wings of the auto-correlation do, however, seem to
extend to large values, with continued structure. Note that the
apparent flat tops of the auto-correlations is due to excising the
zero-lag value, which is contaminated by white noise. After this, the
functions have simply been normalized to 1.

The optical auto-correlations in Figure \ref{autocorr} are all
generally broader than their X-ray counterparts (FWHM$\sim$4\,s). One
would normally expect this from a consideration of the energy
scales: higher-energy emission is produced in an optically thin
medium, and can 
escape easily. The $u'$ curve is consistent with the others from
the night of the 17th, given its lower sampling rate. The optical
auto-correlation functions also show some structure at larger lags,
which we haven't shown as it would make the central peaks hard to
see. 

In Durant et al. (2008) we present the cross-correlation functions
between various energy bands of the RXTE/PCU and the ULTRACAM
observations. We find that there is a strong anti-correlation with the
optical arriving earlier than the X-rays on timescales of 1--10\,s,
followed by a 
much weaker positive response after $\delta t=0$, for the softest X-ray
energies. This is a similar result to the ``precursor'' signal in the
cross-correlation function derived for \xte\ by Kanbach et al (2001).
For medium energies, the correlation function is similar but
narrower, and we find no correlation for high X-ray energies (where
noise dominates). Please refer to this paper for further details.

\subsubsection{Dynamic analysis}

We produced dynamic power spectra
(based on successive small sections of each light curve)
from our rapid timing observations. 

 At first glance, the dynamic power spectra show the same information
as the average power spectra above: power concentrated in the
0.01--0.1\,Hz range, with no abrupt changes thoughout the observation
windows. The two sets
of optical dynamis power spectra taken simultanously look identical. For the
X-ray, the fluctuation with time is
totally consistent with noise. For the optical, however, there is a
hint of structure, but again nothing that can be significantly
distiguished from noise.

\section{Discussion}

A summary of our results:-
\swift\ has not returned to a quiescent state, either in X-rays
(ASM data), gamma-rays (INTEGRAL data) or optical; rather, it seems to
have re-brightened somewhat, compared to one year before our
observations. We do not know the
current state of radio emission, only the upper limits of Soleri et
al. (2008). The optical light curves show superhump-like modulations
with a period of 3.2\,hr, increasingly significant towards the blue,
but not detected in X-rays. The X-ray spectrum is well-described by a
single power law of spectral index $\alpha=1.54$ from 2--60\,keV, with
hints of a possible soft disc component and an emission line near
6\,keV. The optical spectrum shows no sharp features beyond
interstellar ones, only small deviations from a power law.
The SED shows a need for separate components for each wave-band,
in particular, that the optical cannot be a continuation of a flat
synchrotron-dominated spectrum.
In contrast to previous observations, we find no significant QPOs in
power spectra of the X-ray light curves, but we do find $\sim$0.08\,Hz
QPOs in the optical; aside from these, all power spectra are
well-described by two zero-centred Lorentzian functions, with widths
$\sim$0.05\,Hz and $\sim$1.5\,Hz. Autocorrelation functions show a
narrower peak ($\sim$2\,s) for the X-ray emission than the optical
($\sim3$--4\,s), which are consistent with one-another for all the
$u'g'r'$ bands. The cross-correlation functions (Durant et al. 2008)
show the optical leading by 1--10\,s and strongly anti-correlated with
the X-rays, for the lower energies of the RXTE
range. Dynamic power spectra show features which may be short-lived
QPOs in the 0.02-0.1\,Hz region, more poorly defined in the
X-rays than in the optical.

From the work of CB07 and Zurita et al. (2008), we suspect that the
system contains a stellar-mass black hole and M2V-type companion
(which has an undetectable contribution to the current total optical
luminosity of the binary), and is
located at a distance of several kpc, significantly above the Galactic
Plane (in the {\em halo}, since its height of order $>1$\,kpc is much
larger than the disc scale height). Our measurement of interstellar
absorption, particularly of Na, strongly supports a large distance but
also suggests an unknown amount of absorbtion internal to the system.

\subsection{Evolution}
The optical evolution of this source since its 2005 outburst and
discovery is more completely analyzed in Zurita et al. (2008). It is
clear that the compact object in the system, probably a black hole, is
continuing to accrete material, which is powering the observed
luminosity. The system was not observed to leave its low/hard state,
indicating that some low density/high energy region continued to exist
throughout the outburst phase and afterwards. The X-ray spectrum has
changed very little, except for the bolometric luminosity.

There is apparent evolution in the overall shape of the optical
spectrum between CB07, WHT/ISIS and VLT/FORS2 representing two years 
of the system's life since outburst. This is also seen in
the broad-band 
photometry values in Table \ref{photvalues} compared to CB07 - these
are a more accurate measure of the change of spectral slope. The
wiggles in the WHT/ISIS spectrum are not present either earlier (CB07)
or later (out VLT/FORS and NOT/ALFOSC) spectra.

 The smooth continuum and the lack of
lines in maintained throughout, although hydrogen (H$_\alpha$) emission
was seen immediately after the outburst (Torres et al, 2005). The red
part of the spectrum changes very little, and indeed the R-band
magnitude has remained remarkably constant (see Zurita et
al. 2008). The blue part of the spectrum has changed, however: the
large peak at $\sim$4000\AA\, dip at $\sim$4500\AA\ and second peak at
$\sim$5000\AA\ seen in CB07 were not evident in 2006, but had recovered
to some extent by 2007. It would appear that the process which had
caused the enhanced emission during the outburst flare is continuing
at a lower level, and that this had maintained the optical emission
through the two-year period. Ongoing activity is also suggested by the
ASM and INTEGRAL light curves.

In terms of timing, the gross power spectrum has remained qualitatively
similar: flat at low frequencies and flickering at high frequencies. A
QPO was, however, initially detected at 0.6\,Hz by Morgan et
al. (2005) and later at 0.4\,Hz by Ramadevi \& Seetha (2005); an
evolution that was tracked in many RXTE observations by Zhang et al. (2007);
in this work we see such a QPO, relatively weak, but only in the optical
data and not in the X-rays, and at a much lower frequency.

The slowing of characteristic QPO frequencies and time-scales has been
seen for other transient systems (McClintock
\& Remillard, 2006; Shahbaz et al. 2005, 2003), and is commonly taken to
indicate an expanding inner radius of the accretion disc, inside of
which a low density ADAF region forms. The linking of such a QPO
frequency with Keplarian periods may be misleading, however: it is not
obvious how a rotation rate translates into luminosity variations, and
why there would exist preferred orbits within the disc. If for this
system a disc existed at small radii well after the outburst (as
suggested by Miller et al. 2006a), then the frequency characterised by
the QPOs seen cannot have a Keplerian interpretation.

More simple
origins for variability might be disc-corona-jet interaction,
turbulence timescales or magnetic field production/migration
timescales; these might depend on factors such as accretion rate,
magnetic/particle energy balance, jet efficiency, disc density profile
etc. Why these in turn should evolve raises further questions;
recurrent outbursting episodes suggested hysteresis in X-ray
transients (e.g., Malzac, 2007), but the maintenance of a low/hard
state in this case would imply only a small change in the mass flow
rate. The re-brightening of the source since 2006, and its associated
bluer optical colours point to somewhat increasing mass flow from 2006
to 2007.

\subsection{Emission mechanisms}
The X-ray spectrum is typical of Comptonized emission from a
population of energetic particles. This in turn requires a source of
lower energy photons to scatter (possibly seen at the lowest edge of
our X-ray spectrum, and claimed by Miller at al. 2006, and Ramadevi \&
Seetha, 2007) and a replenishable energy reservoir in the
particles. At the same time, we suggested the possibility of a 
contribution of cyclotron emission in the optical, requiring
significant energy content in the magnetic field. Note that
Beloborodov (1999) showed that emission from 
magnetically driven clouds moving away from the disc would not produce
significant re-processing in cooler material, if the bulk motion was
mildly relativistic. 
{ We cannot exclude either a static optically
thick syclotron emitting region nor multiple blackbodies where
particular temperatures are favoured. From the cross-correlation, we
can discount simple reprocessing being an important factor in the
optical emission.}

The case is further complicated by the {\em superhump}
contribution. This accounts for about 10\% of the variation in the
optical region, but 
increasing towards the UV. If it is indeed due to a superhump process
(i.e., either tidal resonance between the disc and orbital periods, or
changing area of the disc; Haswell et al. 2001), then the emission is
thermal and very hot with a black-body-like peak blueward of B. We do not
see any lines from this emission, but we do not have good spectral
sensitivity shortward of 400\,nm. The 2006 WHT spectrum showed less of
a blue component. This could also be the reason that no QPO
was seen in the $u'$-band power spectrum - thermal emission may already be
dominating at these energies. 

Finally there is the case of jet emission. Malzac et al. (2007)
presented a  model for how jet and disc emission could be coupled
in a black hole system such as this. Jet radio emission was certainly seen
earlier, closer to the onset of the outburst (CB07; Fender et
al. 2005), and Soleri et al's limits are not stringent enough to
exclude even a stronger jet than before. Jet/synchrotron emission
cannot, however, be significant in the optical. 

\subsection{Dynamic behaviour}
In Durant et al. (2008) we suggest that the cross-correlation function
implies  that the
optically emitting region was driving rather than responding to the
higher-energy emission. Here we add to this that the optical
auto-correlation functions are {\em broader} than the X-rays, so it
would appear that a relatively slowly building process in the optical
leads to a faster and anti-correlated X-ray response. Furthermore, the
power spectra show that this process occurs as a distribution of small
flares (flickering), with a break at a characteristic time-scale
corresponding to a period of $\sim20$\,s.

Although Malzac et al. (2007) produced a model considering the
dynamic interaction of a jet and accretion disc, in an effort to
explain the optical/X-ray cross-correlation function for \xte, they
explicitly consider only optical emission dominated by synchrotron
emission, with some additional X-ray re-processing by dense disc
material. In our case, the optical is not a simple continuation of a
synchrotron spectrum int he radio (which extrapolates to below the
optical emission); nevertheless, the idea of a
magnetic energy reservoir may still be valid, as it can produce the
types of dynamic behaviour and feedback observed here. We would be
interested to see if their model has a parameter space to match the
details given here. 

\subsection{Comparisons}

It is interesting to note, that \swift\ is the highest Galactic
latitude SXRT after \xte. The latter object shows some of the same
characteristics, which set it apart from the bulk of the SXRT
population: short-period superhump/orbital modulation of 4.1\,hr,
a persistent low/hard state, similar power spectra and unusually bright
optical emission (Hynes et al., 2003; Shahbaz et al. 2005). The X-ray/optical
cross-correlation function for \xte\ was the first to unambiguously
show an optical-leading anti-correlation component (``precognition
peak'', in some sources), although a positive, standard, optical lag
signal was dominant (Kanbach et al. 2001).

Unlike \swift, \xte\ did settle into quiescence.
In this state \xte\ also showed a power spectrum which could be
described as a power law with break or power law
plus QPO. One interpretation of this is to connect this characteristic
frequency with the inner edge of the
accretion disc: asthe disc-ADAF interface increases in radius,
the system decreased in luminosity (Shahbaz et al., 2005). Can the
changes of state really be explained by variable accretion rate alone? 

A leading anticorrelation was seen once for \gx\ in the past (Motch et
al. 1983). Since this initial and unconfirmed measurement, based on a
very short observation window, \gx\ has changed markedly. In
particular, Gandhi et al. (2008) find the cross-correlation is very
different and weak, in a fainter state, although optical and X-ray
spectral characteristics have not changed much.

Assuming that the threshold for an X-ray binary to leave the Low-Hard
State (where the X-ray spectrum is predominantly a power law, rather than
thermal) is  1 percent of the Eddington luminosity (McClintock
\& Remillard, 2006), we
can place a limit on the black hole mass in this system, under several
assumptions. If the bolometric luminosity can be scaled from CB07 to
the peak observed luminosity at the time of outburst, i.e., if the
spectrum remained the same throughout the process, then for a distance
$d$\,kpc we can place the limit 
\begin{eqnarray*}
1\%\times
L_{Edd}&<&6\times10^{37}\left(\frac{d}{6\,\textrm{kpc}}\right)^2\times
3.5 \textrm{\,erg\,cm$^{-2}$s$^{-1}$}\\
M &<& 6.4M_\odot\left(\frac{d}{6\,\textrm{kpc}}\right)^2
\end{eqnarray*}
Zurita et al. (2008) suggest that a stellar-mass black hole
($M\sim3M_\odot$) and M-type star adequately fit the orbital period and
quiescent (i.e., pre-outburst) luminosity of the system. Only by
obtaining a radial-velocity curve and rotational velocity of the donor
star, combined with ellipsoidal light curves can the black hole mass
be concretely established.

\section{Conclusions}
We have conducted optical and X-ray simultaneous spectral and timing
observations of \swift, while the X-ray binary was still in an active
state following its 2005 outburst. We find a superhump-like blue
modulation and remarkably featureless emission in the optical, and
hard power-law 
spectrum in the X-ray, with possible thermal and iron K-line
contributions. The power spectra are similar in X-rays and optical,
flat below 0.05\,Hz and decreasing at higher frequencies, except that
the optical power spectra show $\sim$0.08\,Hz 
QPOs. Dynamic power spectra show these QPOs to be discreet signals
wandering in frequency and persistent on timescales of
$\sim$10\,min. These, together with the respective auto- and
cross-correlation functions and possible cyclotron emission, suggest
that magnetic processes tie together the disc and high-energy
emission.   

Similar features have been seen for two other interesting SXRTs,
\xte\ and \gx. One suggested reason for the difference of
these sources to normal SXRTs is that the accretion disc extends right
in to the inner-most stable orbit.

\medskip
\section*{Acknowledgments}
MD and TS are funded by the Spanish Ministry of Science under the
grant AYA\,2004\,02646 and AYA\,2007\,66887. 
PG is a Fellow of the Japan Society for the Promotion of Science
(JSPS). Based on observations carried out in ESO programmes 079.D-0535,
and 279.D-5021, and during RXTE Cycle 12.
ULTRACAM was designed and built with funding from PPARC (now
STFC), and used as a visiting instrument at ESO Paranal, and RXTE is
operated by NASA. We are grateful for rapid service observations by NOT.
Partially funded by the Spanish MEC under the
Consolider-Ingenio 2010  Program grant  CSD2006-00070: ``First Science
with the GTC''  ({\tt http://www.iac.es/consolider-ingenio-gtc/}).

Thanks to Cadolle-Bel et al. and Paolo Soleri for use of their data in
our SED plots.

\bsp

\label{lastpage}


\begin{thebibliography}{99}
\bibitem{belo}
Beloborodov, A., 1999, ApJ, 510, L123
\bibitem{phots}
Bessell, M., 1995, IAU Symposium, 167, 175
\bibitem{xte}
Bradt, H., Rothschild, R., Swank, J., 1993, A\&AS, 97, 355
\bibitem{flick}
Bruch, A., 1992, A\&A, 266, 237
\bibitem{xrt}
Burrows, D. N., Hill, J. E., Nousek, J. A., Kennea, J. A., Wells, A.,
Osborne, J. P., Abbey, A. F., Beardmore, A., Mukerjee, K., Short,
A. D. T., Chincarini, G., Campana, S., Citterio, O., Moretti, A.,
Pagani, C., Tagliaferri, G., Giommi, P., Capalbi, M., Tamburelli, F.,
Angelini, L., Cusumano, G., Br\"auninger, H. W., Burkert, W., Hartner,
G. D., 2005, SSRv, 120, 165
\bibitem{cad}
Cadolle Bel, M., Rib\'o, M., Rodriguez, J., Chaty, S., Corbel, S.,
Goldwurm, A., Frontera, F., Farinelli, R., D'Avanzo, P., Tarana, A.,
Ubertini, P., Laurent, P., Goldoni, P., Mirabel, I., 2007, ApJ, 659, 549 
\bibitem{xte2}
Chaty, S., Haswell, C., Malzac, J., Hynes, R., Shrader, C., Cui, W.,
2003, MNRAS, 346, 689
\bibitem{adaf}
Czerny, B., R\'oz\.{a}\'ska, A., Janiuk, A., Z\.{y}cki, P.,
2000, NewAR, 44, 439
\bibitem{ultra2}
Dhillon, V., Marsh, T., Stevenson, M, Atkinson, D., Kerry, P.,
et al. 2007, MNRAS, 378, 825
\bibitem{me}
Durant, M., Gandhi, P., Shahbaz, T., Fabian, A., Miller, J., Dhillon,
V., Marsh, T., 2008, ApJL submitted
\bibitem{cyclo}
Fabian, A., Guilbert, P., Motch, C., Ricketts, M., Ilovaisky, S.,
Chevalier, C., A\&A, 111, L9
\bibitem{rad}
Fender, R., Garrington, S., Muxlow, T., 2005, ATel, 558
\bibitem{jets}
Fender, R. \& Belloni, T., 2004, ARA\&A, 42, 317
\bibitem{poshak} 
Gandhi et al., 2008 in preparation
\bibitem{opt}
Halpern, J., 2005, ATel, 549
\bibitem{superhump}
Haswell, C., King, A., Murray, J., Charles, P., 2001, MNRAS, 321, 475
\bibitem{hom}
Homan, J., Belloni, T.,  2005, Ap\&SS, 300, 107
\bibitem{j1118}
Hynes, R., Haswell, C., Cui, W., Shrader, C., O'Brien, K., Chaty, S.,
Skillman, D., Patterson, J., Horne, K., 2003, MNRAS, 345, 292
\bibitem{orbitlag}
Hynes, R., 2005, ASPC, 330, 237
\bibitem{echoes}
Hynes, R., 2006, AIPC, 840, 88
\bibitem{pca}
Jahoda, K., Swank, J. H., Giles, A. B., Stark, M. J., Strohmayer, T.,
Zhang, W., Morgan, E. H., 1996, SPIE, 2808, 59
\bibitem{pcaback}
Jahoda, K., Markwardt, C., Radeva, Y., Rots, A., Stark, M., Swank, J.,
Strohmayer, T., Zhang, W., 2006, ApJS, 163, 401
\bibitem{nat}
Kanbach, G., Straubmeier, C., Spruit, H., Belloni, T., 2001, Nature,
414, 180
\bibitem{isgri}
Lebrun, F., Leray, J. P., Lavocat, P., Cr\'etolle, J., Arqu\'es, M.,
Blondel, C., Bonnin, C., Bou\'ere, A., Cara, C., Chaleil, T., Daly,
F., Desages, F., Dzitko, H., Horeau, B., Laurent, P., Limousin, O.,
Mathy, F., Mauguen, V., Meignier, F., Molini\'e, F., Poindron, E.,
Rouger, M., Sauvageon, A., Tourrette, T., 2003, A\&A, 411, L141
\bibitem{asm}
Levine, A., Bradt, H., Cui, W., Jernigan, J.., Morgan, E., Remillard,
R., Shirey, R., Smith, D., 1996, ApJ, 469, L33
\bibitem{disc}
Liu, B., Taam, R., Meyer-Hofmeister, E., Meyer, F., 2007, ApJ, 671,
695
\bibitem{jetmod}
Malzac, J., Merloni, A., Fabian, A., 2004, MNRAS, 351, 253
\bibitem{adc}
Malzac, J., 2007, MmSAI, 78, 382
\bibitem{xtedist}
McClintock, J. E., Garcia, M. R., Caldwell, N., Falco, E. E.,
Garnavich, P. M., Zhao, P., 2001, ApJ, 551, L147
\bibitem{BH}
McClintock, J., Remillard, R., 2006, Chapter 4 of ``Compact Stellar
X-ray Sources'', eds. W. Lewin and M. van der Klis, CUP
\bibitem{thunder}
Merloni, A. \& Fabian, A., 2001, MNRAS, 328, 958
\bibitem{lowhard}
Meyer-Hofmeister, E., 2004, A\&A, 423, 321
\bibitem{smalldisc}
Miller, J., Homan, J, Miniutti, G., 2006, ApJ, 652, L113
\bibitem{ironlines}
Miller, J., 2007, ARA\&A, 45, 441
\bibitem{qpo}
Morgan, E., Swank, J., Markwardt, C., Gehrels, N., 2005, ATel, 550
\bibitem{UV}
Morris, D., Burrows, D., Racusin, J., Roming, P., Chester, M.,
Verghetta, R., Markwardt, C., Barthelmy, S., 2005, ATel, 552
\bibitem{gx}
Motch, C., Ricketts, M., Page, C., Ilovaisky, S., Chevalier, C.,
1983, A\&A, 119, 171
\bibitem{discover}
Palmer, D., Barthelmey, S., Cummings, J., Gehrels, N., Krimm, H.,
Markwardt, C., Sakamoto, T., Tueller, J., 2005, ATel, 546
\bibitem{rama}
Ramadevi, M., Seetha, S., 2007, MNRAS, 378, 182
\bibitem{review}
Remillard, R. A., McClintock, J. E., 2008, ARA\&A, 44, 49
\bibitem{hexte}
Rothschild, R. E., Blanco, P. R., Gruber, D. E., Heindl, W. A.,
MacDonald, D. R., Marsden, D. C., Pelling, M. R., Wayne, L. R., Hink,
P. L.,  1998, ApJ, 496, 538
\bibitem{huaqu}
Schwope, A., Thomas, H-C., Mantel, K-H., Haefner, R., Staude, A.,
2003, A\&A, 402, 201
\bibitem{tariq1}
Shahbaz, T., Dhillon, V. S., Marsh, T. R., Casares, J., Zurita, C.,
Charles, P. A., Haswell, C. A., Hynes, R. I., 2005, MNRAS, 362, 975
\bibitem{tariq2}
Shahbaz, T., Dhillon, V. S., Marsh, T. R., Zurita, C., Haswell, C. A.,
Charles, P. A., Hynes, R. I., Casares, J., 2003, MNRAS, 346, 1116
\bibitem{sol}
Soleri, P., Altamirano, D., Fender, R., Casella, P., Tudose, V.,
Maitra, D., Mijnands, R., Belloni, T., Miller-Jones, J., Klein-Wolt,
M., van der Klis, M., 2008
\bibitem{gps}
Stil, J., Taylor, A., Dickey, J., Kavars, D., Martin, P., Rothwell,
T., Boothroyd, A., Lockman, F., \& McClure-Griffiths, N.,  2006, AJ, 132, 1158
\bibitem{loc}
Still, M., Roming, P., Brocksopp, C., Markwardt, C., 2005, ATel, 553
\bibitem{xrts}
Tanaka, Y., \& Shibazaki, N., 1996, ARA\&A, 34, 607
\bibitem{halpha}
Torres, M., Steeghs, D., Garcia, M., McClintock, J., Miller, J.,
Jonker, P., Callanan, P., Zhao, P., Huchra, J., U, Vivian, Hutcheson,
C., 2005, ATel, 551
\bibitem{RMS}
Uttley, P. \& McHardy, I., 2001, MNRAS, 312, 880
\bibitem{foam}
Uzdensky, D., Goodman, J.,  2007, MmSAI, 78, 403
\bibitem{anotherreview}
van der Klis, M., in Compact Stellar X-ray Sources. (eds Lewin
W. H. G., van der Klis  M.), Cambridge Univ. Press, p. 39  
\bibitem{swiftqpos}
Zhang, G.-B., Qu, J.-L., Zhang, S., Zhang, C.-M., Zhang, F., Chen, W.,
Song, L.-M., Yang, S.-P., 2007, ApJ, 659, 1511
\bibitem{christina}
Zurita, C., Durant, M., Torres, M., Shahbaz, T., Casares, J., 2008,
ApJ, accepted

\end{thebibliography}
\end{document}